\documentclass[aps,preprint,showkeys,
]{revtex4}
\usepackage{graphicx}
\usepackage{amsfonts}
\usepackage{amssymb}
\usepackage{amsmath}
\usepackage{epsfig}
\usepackage{color}
\usepackage{multirow}
\begin{document}
\title{Intrinsic localized modes for DNLS equation with competing nonlinearities: bifurcations}
\author{G. L. Alfimov$^{1}$, P.A.Korchagin$^1$}
\author{F.K.Abdullaev$^{2}$}
\affiliation{[1] Moscow Institute of Electronic Engineering,
    Zelenograd, Moscow, 124498, Russia}
\affiliation{[2] Physical-Technical Institute of the Uzbekistan Academy of Sciences, Ch.Aytmatov str.2-B,100084,  Tashkent-84, Uzbekistan}

\begin{abstract}
In the paper, we study nonlinear excitations described by DNLS-type equations with so-called competing nonlinearities. These are the nonlinearities that consist of two power terms with coefficients of different sign. 
Equations of this class are of significant interest in modern physics, particularly in applications related to Bose-Einstein condensate (BEC) theory.
A key feature of these models is the presence of two governing parameters: $\alpha$, which characterizes the coupling between lattice sites, and $\gamma$, which quantifies the balance between competing nonlinearities. Our study focuses on intrinsic localized modes (ILMs) -- solutions that exhibit spatial localization over a few lattice sites. The basic example for our study is the cubic-quartic equation that recently has been used to describe 3D BEC cloud in the mean field approximation with Lee-Huang-Yang corrections. Our approach employs numerical continuation from the anti-continuum limit (ACL) where the coupling between the lattice sites is neglected  (the case $\alpha=0$).  We analyze $\alpha$-dependent branches of the basic ILMs and their bifurcations when $\gamma$ varies. Our study shows that all branches of ILMs originated at anti-continuum limit, except a finite number, bifurcate and do not exist for large values of $\alpha$. We present comprehensive tables of bifurcations for the ILMs that involve not more than 3 excited lattice sites. It is shown that the model  supports nonsymmetric ILMs that have no counterparts in ACL. Also we study the branches of ILMs that can be continued unlimitedly when $\alpha\to \infty$ (called here $\infty$-branches). It was found that for any $\gamma$ there are exactly two (up to symmetries) $\infty$-branches. When $\gamma$ grows these branches undergo a sequence of bifurcations that we also describe.  Finally, we compare our results with the results for the
quadratic-cubic equation  and the cubic-quintic equation 
%$1:2$-model and $2:4$-model (aka DNLS-CQ model) 
and found no qualitative difference in (a) tables of bifurcations, (b) presence of solutions without ACL counterpart and (c) scenario of switching of $\infty$-branches.  
\end{abstract}
\keywords{Discrete Nonlinear Schr\"odinger Equation, intrinsic localized modes, anti-continuum limit, numerical continuation}

\date{\today}
%\date{\date{\today}}
\maketitle
%\documentclass{article}
%\usepackage[utf8]{inputenc}
%\usepackage[T2A]{fontenc}
%\usepackage[english, russian]{babel}
%\usepackage{graphicx}
%\usepackage{subcaption}
%\usepackage{float}
%\usepackage{amssymb,amsmath}

%\graphicspath{ {images/} }

\newcommand{\ApA}{A}
\newcommand{\ApB}B

\newcommand{\pk}[1]{\textcolor{red}{#1}}
 
\newcommand{\fa}[1]{\textcolor{blue}{#1}}

\newcommand{\ga}[1]{\textcolor{magenta}{#1}}

\section{Introduction}\label{Sect:Intro}
In this study, we address the lattice equation
\begin{gather}
    i\frac{d\Psi_n}{d t}+C\left(\Psi_{n+1}-2\Psi_n+\Psi_{n-1}\right)+\varkappa\left|\Psi_n\right|^{p-1}\Psi_n-\Gamma\left|\Psi_n\right|^{q-1}\Psi_n=0, \quad n=0,\pm1,\pm2,\ldots.\label{Eq:Gen_pq}
\end{gather}
Here $\varkappa,\Gamma$ are positive constants, $p,q\in \mathbb{N}$ and $q>p$.  Eq.~(\ref{Eq:Gen_pq}) is a generalization of the classical Discrete Nonlinear Schrodinger (DNLS) Equation, 
\begin{gather}
    i\frac{d\Psi_n}{d t}+C\left(\Psi_{n+1}-2\Psi_n+\Psi_{n-1}\right)+\left|\Psi_n\right|^2\Psi_n=0, \quad n=0,\pm1,\pm2,\ldots,\label{Eq:DNLS}
\end{gather}
that has been studied since 50-es due to its numerous physical applications (see, e.g., the  survey \cite{EJ2003} and the book \cite{K2009}). Instead of the cubic nonlinearity of the DNLS equation, the nonlinearity in Eq.(\ref{Eq:Gen_pq}) includes two competing powers of opposite signs (referred to as {\it competing nonlinearity}).  It is known that the ``plus'' term $\sim\left|\Psi_n\right|^{p-1}\Psi_n$ is responsible for localization of the solution and the ``minus'' term $\sim\left|\Psi_n\right|^{q-1}\Psi_n$ causes its spreading. The interplay of these two competing factors results in new features of the model.

The most studied example of equation with competing nonlinearity is the cubic-quintic DNLS equation that corresponds to $p=3$, $q=5$ in Eq.~(\ref{Eq:Gen_pq}),
\begin{gather}
    i\frac{d\Psi_n}{d t}+C\left(\Psi_{n+1}-2\Psi_n+\Psi_{n-1}\right)+\varkappa\left|\Psi_n\right|^2\Psi_n-\Gamma\left|\Psi_n\right|^4\Psi_n=0, \quad n=0,\pm1,\pm2,\ldots.\label{Eq:DNLS-CQ}
\end{gather}
The applications of the cubic-quintic DNLS equation include the nonlinear optics and the theory of Bose-Einstein condensate (BEC).  It is known, that unlike DNLS equation, this equation supports multistability of steady-states \cite{CTCM2006}, mobility of localized excitations for some specific values  of parameters and has other remarkable features \cite{MHM2007,AT2019}.

Recently, two more models of this kind have received wide discussion in the physical literature. Both of them have been used to describe a mixture of two BECs trapped in a deep optical lattice in the presence of quantum fluctuations (QF). The correction to the mean-field energy, given by quantum fluctuations -- the so-called Le-Huang-Yang (LHY) correction, \cite{P15}, depends on the dimensionality of the system. In the case of the quasi-one-dimensional system (a very narrow in the transverse direction trap) the LHY correction is $E_{LHY}\sim n^{3/2}$, where $n$ is the BEC density. Then in the  Gross-Pitaevskii equation, an additional {\it attractive} nonlinear term $\sim |\Psi|\Psi$ appears \cite{ZYZZ2021,KZ2024}. Then, in the tight-binding approximation, the governing discrete equation reads 
\begin{gather}
    i\frac{d\Psi_n}{d t}+C\left(\Psi_{n+1}-2\Psi_n+\Psi_{n-1}\right)+\varkappa\left|\Psi_n\right|\Psi_n-\Gamma\left|\Psi_n\right|^2\Psi_n=0, \quad n=0,\pm1,\pm2,\ldots.\label{Eq:Droplets01}
\end{gather}
This corresponds to  Eq.~(\ref{Eq:Gen_pq}) with $p=2$ and $q=3$ \cite{Droplets01,Droplets02}. If the BEC is loaded in the cigar-type (elongated) trap, then the LHY correction is $E_{LHY}\sim n^{5/2}$ and the additional {\it repulsive} nonlinear term $\sim |\Psi|^3\Psi$ respectively \cite{Edm20,Droplets03,Deb22,Nat22}. This results in the lattice equation 
\begin{gather}
    i\frac{d\Psi_n}{d t}+C\left(\Psi_{n+1}-2\Psi_n+\Psi_{n-1}\right)+\varkappa\left|\Psi_n\right|^2\Psi_n-\Gamma\left|\Psi_n\right|^3\Psi_n=0, \quad n=0,\pm1,\pm2,\ldots,\label{Eq:Droplets02}
\end{gather}
that is also Eq.~(\ref{Eq:Gen_pq}) with $p=3$, $q=4$, \cite{Droplets03}.

In this study, our attention is focused on {\it intrinsic localized modes} (ILM) described by Eq.~(\ref{Eq:Gen_pq}). This term is used to describe strongly localized excitations when only few sites of the lattice are involved. To this end, we impose the boundary conditions 
\begin{gather*}
    \Psi_n\to 0,\quad n\to\pm\infty.%\label{Eq:BoundCond}
\end{gather*}
To be specific, we concentrate on stationary oscillations of the form
%i.e. $\frac{dH}{dt}=\frac{dN}{dt}=0$.
\begin{gather*}
\Psi_n(t)=e^{i\mu t}\psi_n(t),\quad \mu>0.
\end{gather*}
Then $\psi_n(t)$ satisfies the equation
\begin{gather}
    i\frac{d\psi_n}{d t}+C\left(\psi_{n+1}-2\psi_n+\psi_{n-1}\right)-\mu\psi_n+\varkappa\left|\psi_n\right|^{p-1}\psi_n-\Gamma\left|\psi_n\right|^{q-1}\psi_n=0, \quad n=0,\pm1,\ldots.\label{Eq:Gen_psi}
\end{gather}
After rescaling
\begin{gather*}
    \alpha=\frac{C}\mu,\quad \gamma=\frac{\Gamma}{\mu}\cdot\left(\frac{\mu}{\varkappa}\right)^{\frac{q-1}{p-1}},\quad \psi_n=\left(\frac{\mu}{\varkappa}\right)^{\frac{1}{p-1}}u_n,
\end{gather*}
one arrives at the equation
\begin{gather}
    i\frac{du_n}{d t}+\alpha\left(u_{n+1}-2u_n+u_{n-1}\right)- u_n+\left|u_n\right|^{p-1} u_n-\gamma\left|u_n\right|^{q-1} u_n=0, \quad n=0,\pm1,\ldots.\label{Eq:u_gen_pq_t}
\end{gather}
that is the main equation of our study. We call this equation {\it $p:q$-model}. Our findings are illustrated by the example of $3:4$-model (aka cubic-quartic model). However, we repeat the same analysis for $2:3$-model  and $3:5$-model and found no qualitative difference between these cases.

The paper is organized as follows. In Sect.~\ref{Sect:StatStates} we discuss  general properties of the stationary states for Eq.~(\ref{Eq:u_gen_pq_t}), in the limits $\alpha\to0$ and $\alpha\to\infty$. Sect.~\ref{Sect:Bif} is devoted to $\alpha$-dependent branches of these states and their bifurcations,  with special emphasis on their visualization. Sect.~\ref{Sect:Bif-ComNon} contains the main results of our study. There we describe the bifurcations of the basic branches that are common for all the three equations: 3:4-model, 2:3-model and 3:5-model. Sect.~\ref{Sect:Concl} includes summary and discussion of the results. Some details of our numerical method can be found in Appendix \ref{Sect:Numerics}. Appendix \ref{Sect:Tables} contains tables of bifurcations that we refer to at Sect.~\ref{Sect:Bif-ComNon}.

\section{Stationary states}\label{Sect:StatStates}
The steady-states for Eq.~(\ref{Eq:u_gen_pq}) satisfy the equation
\begin{gather}
    \alpha\left(u_{n+1}-2u_n+u_{n-1}\right)- u_n+\left|u_n\right|^{p-1} u_n-\gamma\left|u_n\right|^{q-1} u_n=0, \quad n=0,\pm1,\pm2,\ldots,\label{Eq:u_gen_pq}
\end{gather}
where $u_n$ do not depend on $t$. ILMs correspond to the boundary conditions 
\begin{gather}
    u_n\to 0,\quad n\to\pm\infty.\label{Eq:BoundCond}
\end{gather}
The solutions of Eq.\,(\ref{Eq:u_gen_pq}) are bi-infinite vectors ${\bf u}=(\ldots u_{-1},|u_0,u_1,\ldots)$ (we mark with vertical bar the zero site of the lattice, where this is important). Eq.~(\ref{Eq:u_gen_pq}) is invariant with respect to the following transforms:
\begin{itemize}
    \item the shift by the lattice spacing, $S$, 
    \begin{gather*}
     S:\quad (\ldots,u_{-1},|u_0,u_1,\ldots)\to
        (\ldots,|u_{-1},u_0,u_{1},\ldots);
    \end{gather*}
    \item the phase shift $u_n\to u_ne^{i\theta}$, $\theta\in \mathbb{R}$. In the case $\theta=\pi$ the phase shift becomes the sign-reversing involution,
    \begin{gather*}
        I_\pm:\quad (\ldots,u_{-1},|u_0,u_1,\ldots)\to (\ldots,-u_{-1},|-u_0,-u_{1},\ldots);
\end{gather*}
    \item  the flip transform centered {\it on site},
\begin{gather*}
I_{\rm on}^+:\quad (\ldots,u_{-1},|u_0,u_1,\ldots)\to (\ldots,u_{1},|u_0,u_{-1},\ldots);
\end{gather*}
     \item  the flip transform centered {\it inter sites},
\begin{gather*}
I_{\rm in}^+:\quad (\ldots,u_{-1},|u_0,u_1,u_2,\ldots)\to (\ldots,u_{1},u_0,|u_{-1},u_{-2},\ldots).
\end{gather*}
\end{itemize}
Evidently, $I_\pm^2=(I_{\rm on}^+)^2=(I_{\rm in}^+)^2=E$ where $E$ is the identity map, so $I_\pm$, $I_{\rm in}^+$ and $I_{\rm on}^+$ are involutions.
%It is straightforward to check that $I_{\rm on}^+I_{\rm in}^+=S$. 
Two more involutions are
\begin{gather*}
I_{\rm on}^-=I_\pm I_{\rm on}^+,\quad I_{\rm in}^-=I_\pm I_{\rm in}^+.
\end{gather*}
We say that a solution ${\bf u}$ of Eq.\,(\ref{Eq:u_gen_pq}) is {\it $I_{\rm on}^+$-symmetric} or {\it $I_{\rm on}^-$-symmetric} if it is invariant with respect to the involution $I_{\rm on}^+$ or $I_{\rm on}^-$. Similarly we define {\it $I_{\rm in}^+$-symmetric} and {\it $I_{\rm in}^-$-symmetric} solutions. 

The quantity
\begin{gather}
J=\overline{u}_nu_{n+1}-u_n\overline{u}_{n+1}
\end{gather}
(the bar above means the complex conjugation) does not depend on $n$. Due to (\ref{Eq:BoundCond}), $J=0$ for ILMs. This implies that either $u_n=0$ (and, therefore, $u_{n+1}=-u_{n-1}$), or
\begin{gather*}
\frac{u_{n+1}}{u_{n}}=\frac{\overline{u}_{n+1}}{\overline{u}_{n}},
\end{gather*}
and the arguments of $u_{n+1}$ and $u_n$ are equal modulo $\pi$.  Due to the phase shift invariance, without loss of generality  we can assume that $u_n\in\mathbb{R}$ for any $n\in\mathbb{Z}$.

\subsection{Anticontinuum limit}\label{Sect:AL}
The case $\alpha=0$ corresponds to the so-called {\it anticontinuum limit} (ACL) \cite{A1997}. In ACL the sites of the lattice become decoupled and the amplitude of each of them satisfies the equation
\begin{gather}
    u(1-|u|^{p-1}+\gamma|u|^{q-1})=0.\label{Eq:Roots}
\end{gather}
Simple algebra shows that Eq.~(\ref{Eq:Roots}) has
\begin{itemize}
    \item[(a)] three roots $u=0$, $u=\pm 1$ if $\gamma=0$;
    \item[(b)] five roots $u=0$, $u=\pm a$, $u=\pm A$, $0<a<A$, if 
    \begin{gather}
        0<\gamma<\gamma^{**}\equiv\left(\frac{p-1}{q-1}\right)\cdot\left(\frac{q-p}{q-1}\right)^{\frac{q-p}{p-1}};\label{Eq:BoundGamma}
    \end{gather}
    \item[(c)] three roots $u=0$ and 
    \begin{gather*}
        u=\pm\left(\frac{q-1}{q-p}\right)^{1/(p-1)}
    \end{gather*}
    if $\gamma=\gamma^{**}$;
    \item[(d)] one root $u=0$ if $\gamma>\gamma^{**}$.
\end{itemize}
In the case (d) no ILM solution exists for any $\alpha>0$. The cases (a) and (c) are not the cases of general situation (note that the case (a) corresponds to DNLS equation if $p=3$). So, in this study we focus mainly on the case (b), i.e., we assume that $\gamma$ satisfies the condition (\ref{Eq:BoundGamma}). 

\begin{figure}
    \centerline{\includegraphics[width=0.55\textwidth]{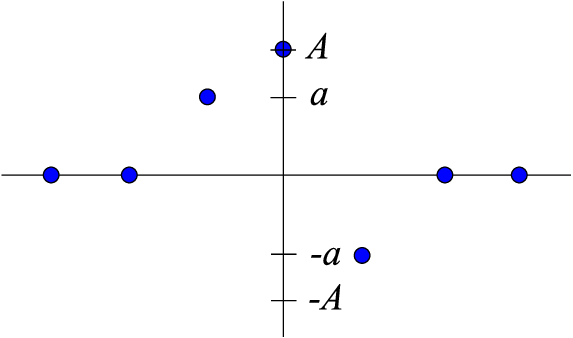}}
    \caption{The mode with the code $\mathcal{A}=(\ldots,00a_+|A_+a_-00\ldots)$}
    \label{Fig:ModeCQ}
\end{figure}

We label the solutions in the ACL by sequences of symbols of some alphabet. If $\gamma=0$ an alphabet of three symbols is sufficient. For the sake of unification, we use the symbols ``$0$'', ``$a_+$'' and ``$a_-$'' that correspond to the amplitudes of the sites $0$, $1$ and $-1$ respectively. In the case (b) we use the alphabet of five symbols, ``$0$'', ``$a_+$'', ``$a_-$'', ``$A_+$'', ``$A_-$''. They correspond to the amplitudes $0$, $a$, $-a$, $A$ or $-A$ of the sites respectively. When coding, we also mark zero position by vertical bar. For instance, the coding sequence
\begin{gather*}
    {\mathcal A}=(\ldots00a_+|A_+a_-00\ldots)
\end{gather*}
denotes the solution with amplitudes $u_0=A$, $u_1=-a$, $u_{-1}=a$  and $u_n=0$ for $n>1$ and $n<-1$ (see Fig.\,\ref{Fig:ModeCQ}). Note that the codes of ILMs start and end with infinite number of symbols ``$0$'', so that only finite number of symbols are nonzero. To shorten the notations, we cut off infinite number of zero symbols left the first nonzero symbol and right to the last nonzero symbol keeping only finite number of entries (for instance, the truncation of the code $\mathcal A$ is $(a_+|A_+a_-)$). We call {\it the length of the code} the number of symbols in the truncated code. For instance, the code $\mathcal A$ is of the length 3.

\subsection{Continuum limit}\label{Sect:ContLimit}

When $\alpha$ is large one can approximate the lattice solution ${\bf u}=\{u_n\}$ of Eq.~(\ref{Eq:u_gen_pq}) by continuous solution $u(x)$ of the equation
\begin{gather}
\frac{d^2U}{dx^2}-U+|U|^{p-1}U-\gamma|U|^{q-1}U=0\label{Eq:ContEq}
\end{gather}
where $u_n=U(n/\sqrt{\alpha})+o(1/\sqrt{\alpha})$. Due to  (\ref{Eq:BoundCond}), the boundary conditions for $U(x)$ are
\begin{gather}
U(x)\to 0,\quad x\to\pm \infty.\label{Bound_U}
\end{gather}
\begin{figure}
    \centerline{\includegraphics[width=0.85\textwidth]{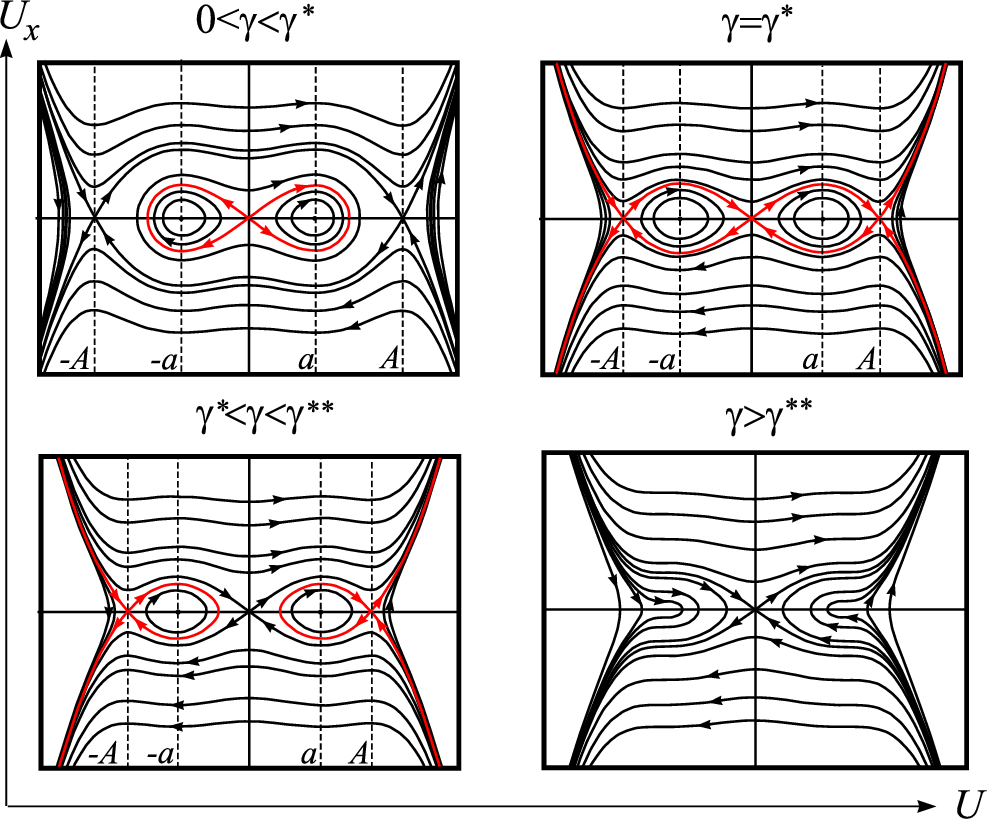}}
    \caption{Typical phase portraits for Eq.~(\ref{Eq:ContEq}) for various values of $\gamma$. Homoclinic separatices exist for $0<\gamma<\gamma^*$ where $\gamma^*$ is given by formula (\ref{gamma*}) }
    \label{Fig:PhPort}
\end{figure}
If $\gamma=0$ the solutions of (\ref{Eq:ContEq})-(\ref{Bound_U}) are
\begin{gather*}
U(x)=\pm \left[\left(\frac{p+1}2\right){\rm sech}^2\left(\frac{(p-1) x}2\right)\right]^{1/(p-1)}.
\end{gather*}
If $0<\gamma<\gamma^{**}$ the equation (\ref{Eq:ContEq}) has five equilibrium states $U=0$, $U=\pm a$, $U=\pm A$. The typical phase portraits for Eq.~(\ref{Eq:ContEq}) is shown in Fig.\ref{Fig:PhPort}. The solution of (\ref{Eq:ContEq})-(\ref{Bound_U}) corresponds to homoclinic separatices of the equilibrium $(0,0)$. They exist for $0<\gamma<\gamma^*$ where 
\begin{gather}
0<\gamma<\gamma^*\equiv\frac{(p-1)(q+1)(2q-2p)^{\frac{q-p}{p-1}}}{\left((q-1)(p+1)\right))^{\frac{q-p}{p-1}}}.\label{gamma*}
\end{gather}
Generically, the solution $U(x)$ cannot be expressed in   explicit  form. Some exact solutions for particular cases $p=2$, $q=3$ and $p=3$, $q=5$ are given in Table \ref{Tab:DataModels}.

\subsection{Stability}\label{Sect:Stability}

The stability of a stationary state ${\bf u}=\{u_n\}$ is determined by the eigenvalue problem, \cite{K2009},
\begin{gather}
\omega\left(\begin{array}{c}
{\bf v}\\{\bf w}
\end{array}
\right)=
\left(\begin{array}{cc}
{\bf 0}& {-\bf L_-}\\
{\bf L_+}& {\bf 0}
\end{array}
\right)
\left(\begin{array}{c}
{\bf v}\\{\bf w}
\end{array}
\right).\label{Eq:EigenProb}
\end{gather}
Here ${\bf v}=\{v_n\}$, ${\bf w}=\{v_n\}$ are  bi-infinite vectors. The infinite matrices  ${\bf L_\pm}$ are
%\small
\begin{gather*}
{\bf L_-}=\left(
\begin{array}{ccccc}
\ddots&   \vdots    &   \vdots    &    \vdots    &  \vdots    \\[2mm]
\cdots&       D^-_{n-1} &     -\alpha         &   0             &    \cdots\\[2mm]
\cdots&                     -\alpha&    D^-_{n} &               -\alpha&    \ldots\\[2mm]
\cdots&                       0&              -\alpha&D^-_{n+1} &     \ldots\\[2mm]
\cdots&    \vdots   &       \vdots&         \vdots&        \ddots
\end{array}
\right),~
{\bf L_+}=\left(
\begin{array}{ccccc}
\ddots&   \vdots    &   \vdots    &    \vdots    &  \vdots    \\[2mm]
\cdots&       D^+_{n-1} &     -\alpha         &   0             &    \cdots\\[2mm]
\cdots&                     -\alpha&    D^+_{n} &               -\alpha&    \ldots\\[2mm]
\cdots&                       0&              -\alpha&D^+_{n+1} &     \ldots\\[2mm]
\cdots&    \vdots   &       \vdots&         \vdots&        \ddots
\end{array}
\right)
\end{gather*}
\normalsize
where
\begin{gather*}
D_n^-=2\alpha+1-|u_n|^{p-1}+\gamma |u_n|^{q-1},\quad D_n^+=2\alpha+1-p|u_n|^{p-1}+q|u_n|^{q-1}.
\end{gather*}
%where $(\ldots,u_{-1},u_0,u_{1},\ldots)$ is the ILM.
If $\omega$ is an eigenvalue of (\ref{Eq:EigenProb}) then $-\omega$, $\omega^*$ and $-\omega^*$ are also eigenvalues of (\ref{Eq:EigenProb}). We say that a stationary mode ${\bf u}=\{u_n\}$ is {\it linearly stable} if ${\rm Re}~\omega=0$ for all the eigenvalues of (\ref{Eq:EigenProb}).  Since 
\begin{gather*}
{\bf L}_-{\bf u}={\bf 0},
\end{gather*}
the operator ${\bf L}_-$ is degenerate and $\omega=0$ belongs to the spectrum of (\ref{Eq:EigenProb}). Alternatively, one can rewrite (\ref{Eq:EigenProb}) in the form, 
\begin{gather*}
\omega {\bf v}=-{\bf L}_-{\bf w},\quad \omega {\bf w}={\bf L}_+{\bf v}.
\end{gather*}
Therefore
\begin{gather}
{\bf L}_+{\bf L}_-{\bf w}=\lambda {\bf w},\quad \lambda=-\omega^2.\label{Eq:EigenProb02}
\end{gather}
and we say that ILM is stable if $\lambda\geq 0$ for all the eigenvalues of (\ref{Eq:EigenProb02}).

\section{Branches of the ILMs: some general comments}\label{Sect:Bif}

It is known, that each ILMs can be continued from the ACL to some interval $0<\alpha<\alpha_c$ \cite{A1997,MA1994}. While this continuation, the nonlinear mode remains localized and the asymptotic behavior of  its ``tails'' as $n\to\pm\infty$ is exponential.   We label $\alpha$-depended branches of ILMs originated in the ACL by the same code that has the solution at $\alpha=0$. In what follows we assume that the modes have codes of finite length.

When $\alpha$ grows the branches may undergo bifurcations. Two of the bifurcations are essential:
\begin{itemize}
\item At some  $\alpha=\alpha_{c}$ a branch of ILMs undergoes  {\it the fold} bifurcation and disappears by merging with another branch of ILMs. 
\item  At some $\alpha=\alpha_{c}$ a branch of $I_{\rm on}^\pm$- or $I_{\rm in}^\pm$-symmetric ILMs undergoes {\it the pitchfork} bifurcation giving birth to a pair of branches of non-symmetric ILMs related to each other by $I_{\rm on}^\pm$- (respectively, $I_{\rm in}^\pm$-) involution.
\end{itemize}

Below we show that the ``most part'' of the branches of ILMs undergo some bifurcation when $\alpha$ grows and disappear. Only few branches  can be continued from ACL ($\alpha=0$) to the continuum limit $\alpha\to \infty$. We call them {\it  $\infty$-branches.}

\subsection{Visual representation} 

Let $\gamma\geq 0$ and $\alpha>0$ be fixed. Then any pair $(u_0^*,u_1^*)$ determines uniquely the solution of Eq.~(\ref{Eq:u_gen_pq}),  bi-infinite vector ${\bf u}^*=(\ldots,u_{-1}^*|,u_0^*,u_1^*,\ldots)$,  by the recurrence relation
\begin{gather*}
    u_{n+1}^*=2u_n^*-u_{n-1}^*- \frac1\alpha \left(u_n^*-\left|u_n^*\right|^{p-1} u_n^*+\gamma\left|u_n^*\right|^{q-1} u_n^*\right), \quad n=0,\pm1,\pm2,\ldots .
\end{gather*}
However it may not be ILM, since it may not satisfy the boundary conditions $u_n\to 0$ when $n\to \pm\infty$. 

Consider the plane $(u_0,u_1)$. In order ${\bf u}^*$ to be an ILM, the point $(u_0^*,u_1^*)$ has to lie on $C^1$-smooth stable manifold of the equilibrium $(0,0)$ (see e.g. \cite{CTCM2006}). This means that the pair $(u_0^*,u_1^*)$ has to satisfy some relation of the form $F_s(u_0^*,u_1^*)=0$, where $F_s$ is a $C^1$-smooth function of two arguments. Simultaneously, $(u_0^*,u_1^*)$  has to lie on $C^1$-smooth unstable manifold of the equilibrium $(0,0)$, i.e, another relation of the form $F_u(u_0^*,u_1^*)=0$ must be satisfied ($F_u$ is also a $C^1$-smooth function of two arguments).  So, for generic values $\gamma$ and $\alpha$, ILMs correspond to {\it discrete} set of points $(u_0^*,u_1^*)$ in the plane $(u_0,u_1)$ that are intersections of stable and unstable manifolds of the equilibrium $(0,0)$. Note that if ${\bf u}^*$ is an ILM, then due to the shift invariance any pair $(u_m^*,u_{m+1}^*)$, $m\in\mathbb{Z}$, belongs to this discrete set.

Let now $\gamma\geq0$ be fixed and $\alpha>0$ varies. The functions $F_s$ and $F_u$ become $C^1$-smooth functions of $u_0$, $u_1$ and $\alpha$. The $\alpha$-dependent branches of ILMs originated at ACL are described by the system
\begin{gather*}
\left\{
\begin{array}{l}
F_s(u_0,u_1,\alpha)=0,\\[1mm]
F_u(u_0,u_1,\alpha)=0.
\end{array}
\right.
\end{gather*}
In 3D space $(u_0,u_1,\alpha)$ they correspond to $C^1$-smooth curves. These curves start at 25 points of the plane $\alpha=0$: $(0,0,0)$, $(\pm a,\pm a,0)$, $(\pm a,\pm A,0)$, $(\pm A,\pm a,0)$, $(\pm A,\pm A,0)$ (all the combinations of the signs plus and minus are possible). The continuation along a curve $\Gamma$ is possible at any {\it regular} point $(u_0^*,u_1^*,\alpha^*)\in \Gamma$, i.e. at the point where the matrix
\begin{gather*}
DF(u_0^*,u_1^*,\alpha^*)=\left(
\begin{array}{lll}
\displaystyle{\frac{\partial F_s}{\partial u_0}(u_0^*,u_1^*,\alpha^*)}~ &
\displaystyle{\frac{\partial F_s}{\partial u_1}(u_0^*,u_1^*,\alpha^*)}~ &
\displaystyle{\frac{\partial F_s}{\partial \alpha}(u_0^*,u_1^*,\alpha^*)} \\[2mm]
\displaystyle{\frac{\partial F_u}{\partial u_0}(u_0^*,u_1^*,\alpha^*)}~ &
\displaystyle{\frac{\partial F_u}{\partial u_1}(u_0^*,u_1^*,\alpha^*)}~ &
\displaystyle{\frac{\partial F_u}{\partial \alpha}(u_0^*,u_1^*,\alpha^*)}
\end{array}
\right)  
\end{gather*}
has rank 2. The curves cannot intersect each other at a regular point.
A curve that consists of regular points only, we call {\it a regular curve} and the corresponding branch of ILMs {\it a regular branch}. 

In 3D space $(u_0,u_1,\alpha)$ the involutions $I_{\rm in}^+$ and $I_{\rm in}^-$ correspond to the mirror symmetries with respect to the planes $R_+:(u_0=u_1)$ and $R_-:(u_0=-u_1)$. The curves that correspond to the branches of $I_{\rm in}^+$-symmetric (or $I_{\rm in}^-$-symmetric) ILMs lie entirely in the planes $R_+$ (respectively, $R_-$). Note that the rank of the matrix $DF$ is less than two at any point of such a curve.

When two branches of ILMs undergo a fold bifurcation at some $\alpha=\alpha_{cr}$, their curves merge in 3D space $(u_0,u_1,\alpha)$. In this case, we say that the branches (and the corresponding curves) form {\it a loop}. If all points of a loop are regular, we call it {\it a regular loop}. A regular loop cannot intersect either $R_+$ or $R_-$. If a loop intersects $R_+$ (or $R_-$), the point of intersection is not regular and corresponds to the pitchfork bifurcation. In this case, the loop is formed by two curves related to each other by mirror symmetry with respect to $R_+$ (or $R_-$). These curves adjoin the curve that lies in $R_+$ (or $R_-$) and corresponds to a branch of $I_{\rm in}^+$-symmetric (or $I_{\rm in}^-$-symmetric) ILMs. 

\begin{figure}
    \centerline{\includegraphics[width=0.75\textwidth]{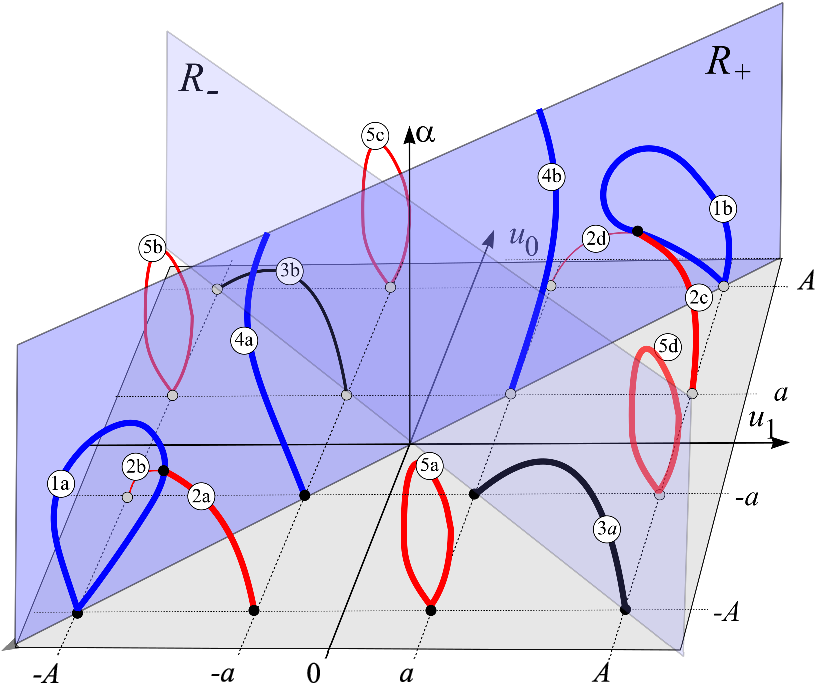}}
    \caption{A hypothetical 3D view of some branches of ILMs in coordinates $(u_0,u_1,\alpha)$ when $\gamma$ is fixed. 1a and 1b: the loops that appear due to fold bifurcations of pairs of $I_{\rm in}^+$-symmetric ILMs. The loops 1a and 1b are $I_{\rm in}^-$-symmetric to each other. The codes of ILMs for the loop 1a are of the form $(\ldots,|A_-A_-,\ldots)$ whereas ones for 1b are of the form  $(\ldots,|A_+A_+,\ldots)$. 2a and 2b: the curves that correspond to two $I_{\rm in}^+$-symmetric branches that undergo pitchfork bifurcation, adjoining to the loop 1a. 2c and 2d: the same, but adjoining to the loop 1b. 3a: the  loop formed by two $I_{\rm in}^-$-symmetric branches, one having the code of the form $(\ldots |a_-a_+\ldots)$ and another of the form $(\ldots |A_-A_+\ldots)$. 3b: its $I_{\rm in}^+$-symmetric counterpart.     
    4a and 4b:  the curves corresponding to the branches of ILMs with the codes of the form $(\ldots,|a_\pm,a_\pm,\ldots)$. They do not undergo any bifurcation. 5a: the regular loop that arise due to the fold bifurcation of nonsymmetric ILM branches. The codes of both the branches are $(\ldots|A_-a_+\ldots)$. 5b,5c,5d: the regular loops that are related to 5a by means  of  $I_{\rm in}^-$, $I_{\rm in}^+$ and $I_{\rm in}^-I_{\rm in}^+$ involutions.
    }
    \label{Fig:3D_general}
\end{figure}

Conceptual sketch of possible position of branches of ILMs in 3D space $(u_0,u_1,\alpha)$ is presented in Fig.~\ref{Fig:3D_general}. 

Alternatively, one can consider the branches of ILMs in 3D space $(u_0,p_0,\alpha)$ where $p_0=u_1-u_{-1}$. Then the involutions $I_{\rm on}^+$ and $I_{\rm on}^-$ correspond to mirror symmetries with respect to the planes $Q_+:(p_0=0)$ and $Q_-:(u_0=0)$. The corresponding curves start at 35 points of the plane $(u_0,p_0,0)$, specifically $(s,\pm A)$,$(s,\pm(A-a))$, $(s,\pm a)$, $(s,0)$, where $s\in\{0,\pm a,\pm A\}$ (all combinations of signs plus and minus are admissible). The curves that correspond to branches of $I_{\rm on}^+$-symmetric (or $I_{\rm on}^-$-symmetric) ILMs lie entirely in the planes $Q_+$ (or, respectively, $Q_-$). Again, if all the points of a curve are regular, the curve does not intersect either $Q_+$ or $Q_-$.

The 3D graphical visualization has proven to be an efficient tool to describe the branches  of ILMs and their reconnections that occur when $\gamma$ varies.

\subsection{Restrictions for the regular loops}\label{Eq:Restrictions}

Let two regular branches with codes ${\mathcal A}_1$ and ${\mathcal A}_2$ merge forming a regular loop. 3D representation of the loop neither intersect the planes $R_\pm$ in the space $(u_0,u_1,\alpha)$ nor the planes $Q_\pm$ in the space $(u_0,p_0,\alpha)$. This fact imposes restrictions to the codes ${\mathcal A}_1$ and ${\mathcal A}_2$.

\begin{figure}
    \centerline{\includegraphics[width=0.85\textwidth]{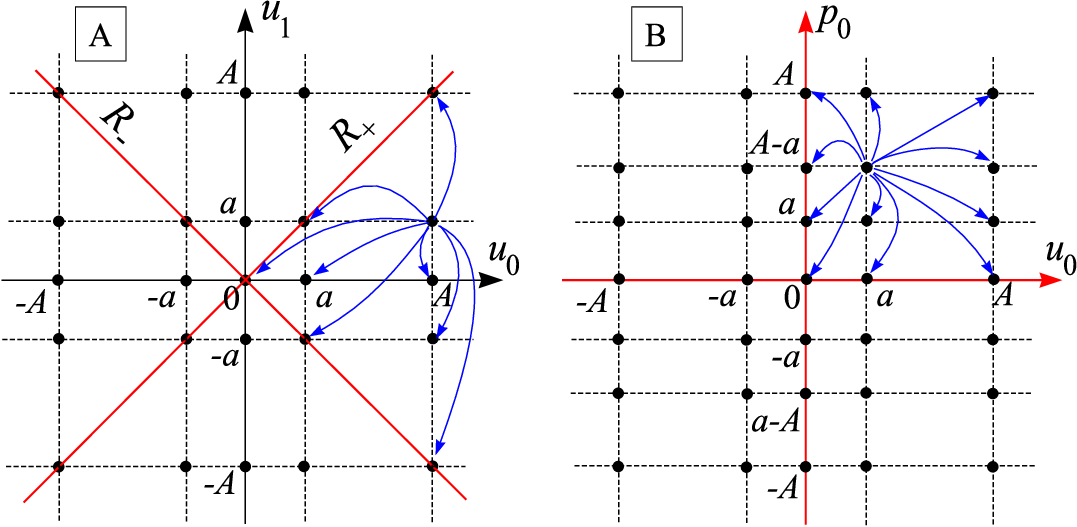}}
    \caption{The planes $(u_0,u_1)$ (panel A) and $(u_0,p_0)$ (panel B). Panel A: a regular branch of ILMs cannot intersect the lines $u_0=u_1$ and $u_0=-u_1$. If a regular branch has the code of the form ${\mathcal A}_1=(\ldots A_+a_+\ldots)$, then there are 9 possibilities for pair of corresponding symbols at another branch ${\mathcal A}_2$ to merge with. Panel B: a regular branch of ILMs cannot intersect the lines $u_0=0$ and $p_0=0$. The possibilities for triples of symbols in  ${\mathcal A}_2$ if ${\mathcal A}_1=(\ldots a_+a_+A_+\ldots)$ are shown. }
    \label{Fig:2_planes}
\end{figure}

We illustrate these restrictions by the following example. Let the code ${\mathcal A}_1$ include the fragment $(\ldots A_+a_+A_-\ldots)$. Due to the shift invariance one may treat each of these symbols as a symbol on zeroth place. Two subsequent symbols $(A_+a_+)$ of ${\mathcal A}_1$ must correspond to one of the following pairs in ${\mathcal A}_2$: $(A_+a_+)$, $(A_+A_+)$, $(A_+0)$, $(A_+a_-)$, $(A_+A_-)$, $(a_+a_+)$, $(a_+0)$, $(a_+a_-)$, $(0,0)$, since otherwise the loop intersects one of the planes $R_\pm$ in the space $(u_0,u_1,\alpha)$ (see Fig.~\ref{Fig:2_planes}). Then the possibilities  $(A_+a_-)$, $(A_+A_-)$ and $(a_+a_-)$ should be also discarded, since in these cases the intersections with the planes $Q_\pm$ in the space $(u_0,p_0,\alpha)$ occur. Repeating the same procedure for the next two symbols $(a_+A_-)$ we find the following 14 triplets that might correspond to the given fragment,
\begin{align*}
   & (A_+a_+A_-),~ (A_+a_+a_-),~ (A_+A_+A_-),~ (A_+0A_-),~ (A_+0a_-),~ (a_+a_+a_-),~\\[2mm] 
   & (a_+a_+A_-),~ (a_+0A_-),~ (a_+0a_-),~(a_+0,0),~(00A_-),~(00a_-),~(A_+0,0),~ (000).
\end{align*}
Empirically, it was discovered that the symbols $A_+$ and $A_-$ in ${\mathcal A}_1$ never matches to the symbol $0$ in ${\mathcal A}_2$. Assuming this fact, last 5 possibilities also can be discarded. It was found (see Appendix \ref{Sect:Tables}, Table \ref{Tab:Rest}) that the branch of ILMs with the code $(|A_+a_+A_-)$ can merge either with the branch $(a_+|A_+a_+A_-)$ or with the branch $(|A_+a_+a_-)$, depending of the value of $\gamma$. This fact agrees with the reasoning above. 

\subsection{The case of DNLS equation, $\gamma=0$, $p=3$}\label{Sect:Bif-DNLS}

\begin{figure}
    \centerline{\includegraphics[width=0.85\textwidth]{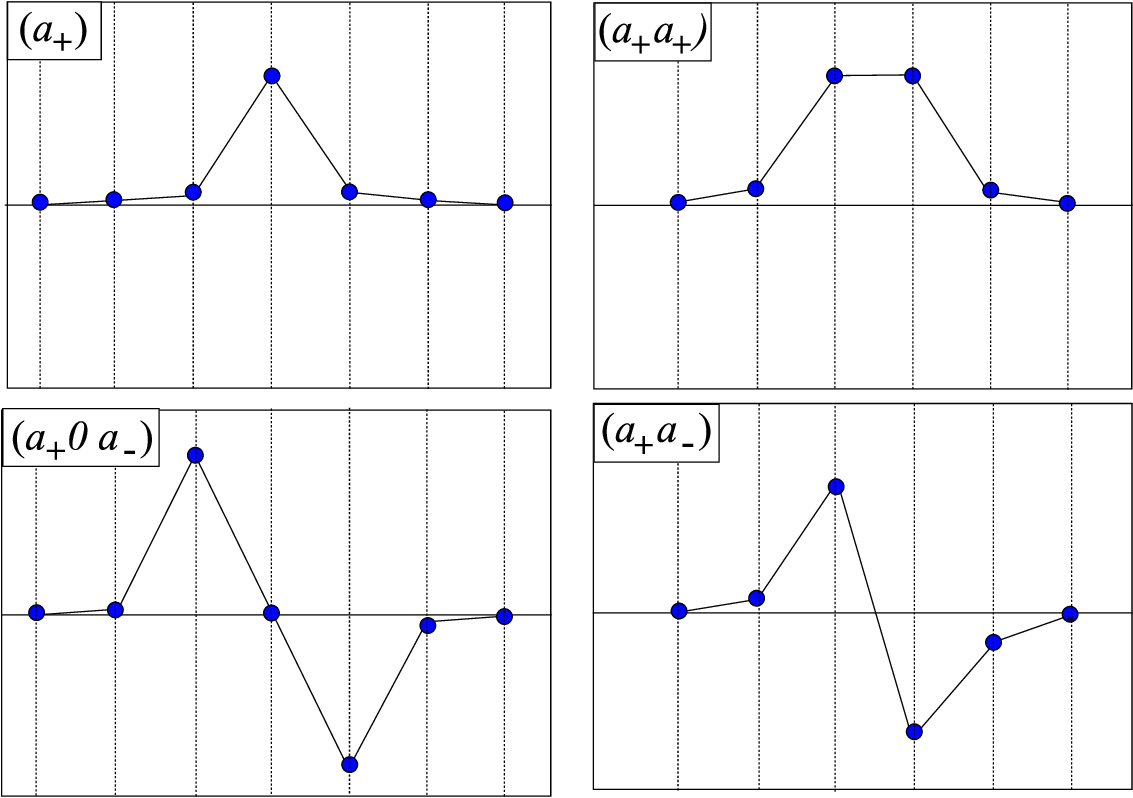}}
    \caption{Basic modes for DNLS equation. $(a_+)$ - the Sievers-Takeno mode, $(a_+a_+)$ - the Page mode, $(a_+a_-)$ and $(a_+0a_-)$ - the twisted modes}
    \label{Fig:Basic_shapes}
\end{figure}

%In order to have a reference when considering %the general case,
%in the general case, Eq.~(\ref{Eq:u_gen_pq}) with $\gamma\ne 0$, 
For the sake of completeness, let us briefly review the results for the most studied particular case of Eq.~(\ref{Eq:u_gen_pq}), the DNLS equation that corresponds to $\gamma=0$ and $p=3$.

1. There are several {\it basic ILMs} that have been discussed in numerous studies (see, e.g., \cite{K2009,ABK04,DKL98}. They are the {\it Sievers-Takeno mode} (the code $(a_+)$), the {\it Page mode} (the code $(a_+a_+)$) and two types of {\it twisted modes}, (the codes $(a_-a_+)$ and $(a_-0a_+)$. The modes above are defined up to the involutions $I_{\rm on}^\pm$, $I_{\rm in}^\pm$. Due to the shift invariance they can be centered at any lattice site.  
The following comments are in order.

(i) The branch of the Sievers-Takeno modes is an $\infty$-branch, i.e. it extends from ACL ($\alpha=0$) to the continuum limit ($\alpha\to\infty$).  The Sievers-Takeno modes are stable.

(ii) The branch of the Page modes is also an $\infty$-branch. The Page modes are unstable.

(iii) The branch of the twisted modes $(a_-a_+)$ exists for $0\leq \alpha<\alpha_{1}$, $\alpha_{1}\approx 0.772$. At $\alpha_1$ it undergoes the fold bifurcation and merge with the branch $(a_-a_-a_+a_+)$. It is stable for $\alpha\in[0;\alpha_{1}^{o})$ where $\alpha_{1}^{o}<\alpha_{1}$ is some threshold value.

(iv) The branch of the twisted modes $(a_-0a_+)$ exists for $0\leq \alpha<\alpha_{2}$, $\alpha_{2}\approx 1.091$. At $\alpha_1$ it undergoes the fold bifurcation and merge with the branch $(a_-a_-0a_+a_+)$. It is also stable for $\alpha\in[0;\alpha_{2}^{o})$ where $\alpha_{2}^{o}<\alpha_{2}$ is some threshold value.

Schematically, the shapes of basic modes are shown in Fig.\ref{Fig:Basic_shapes}. 

\begin{figure}
    \centerline{\includegraphics[width=0.95\textwidth]{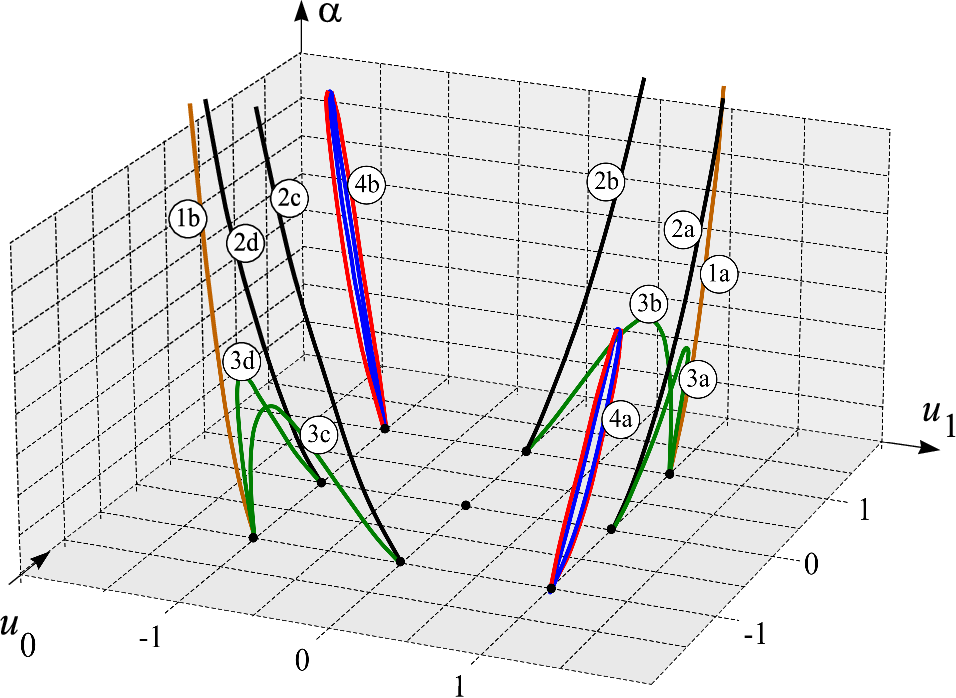}}
    \caption{3D view of some branches of ILMs for DNLS equation. 1a and 1b: the codes $(|a_+a_+)$ and $(|a_-a_-)$; 2a, 2b, 2c and 2d: the codes $(|0 a_+)$, $(|a_+0)$, $(|a_-0)$, $(|0 a_-)$; 3a, 3b, 3c and 3d: the loops that arise due to merging of pairs $(a_+ |0 a_+)$ with $(a_+|a_+a_+)$, $(|a_+0a_+)$ with  $(|a_+a_+a_+)$, $(|a_-0a_-)$ with  $(|a_-a_-a_-)$ and $(a_- |0 a_-)$ with $(a_-|a_-a_-)$ respectively. 4a: the symmetric loop formed by the branches $(|a_-a_+)$ and $(a_-|a_-a_+a_+)$ (blue curve) together with $(a_-|a_-a_+)$ and $(|a_-a_+a_+)$ (red curves) that adjoin to this loop as a result of pitchfork bifurcation; 4b: the same as 4a with  $(|a_+a_-)$ and $(a_+|a_+a_-a_-)$ (the loop), $(a_+|a_+a_-)$ and $(|a_+a_-a_-)$ (the secondary branches).}
    \label{Fig:DNLS}
\end{figure}

2. It was proved by K.Yoshimura (see \cite{Yo16}) that the continuation of any ILM from the ACL is possible for $0\leq\alpha<0.2056$. Numerical computation, \cite{ABK04}, yields the estimation $0\leq\alpha<0.2896$ for the interval where no bifurcation of ILMs occurs.

3. As $\alpha$ grows, no bifurcations of birth of {\it new} ILMs occur. To our best knowledge, this statement has not been proved rigorously but it was supported by numerics in \cite{ABK04}.

4. When $\alpha$ grows, all the ILMs die in bifurcations, except the two, up to the sign: $(a_+)$ (the Sievers-Takeno mode) and $(a_+a_+)$ (the Page mode). Both the modes $(a_+)$ and $(a_+a_+)$  approach the continuous limit as $\alpha\to\infty$, i.e. the solution $U(x)=\sqrt{2}\,{\rm sech}\,x$ of (\ref{Eq:ContEq}). In Fig.~\ref{Fig:DNLS}  some branches of ILMs and their bifurcations are shown in 3D space $(u_0,u_1,\alpha)$.

5. When moving along the branch, ILMs can gain or lose stability. Although the overall picture of stability/instability of ILMs is quite complex, the question of their stability for small values of $\alpha$ has been completely resolved \cite{PKF05}. It has been proved that the stability is completely determined by the code of the branch. The stability of a branch with a code ${\mathcal{A}}$ can be established as follows:
\begin{itemize}
    \item delete all ``$0$'' symbols from the code ${\mathcal{A}}$;
    \item check that in the resulting code the symbols $a_-$ and $a_+$  alternate.
\end{itemize}
Then ILMs of the branch with code ${\mathcal{A}}$ is stable for $\alpha$ small enough. Otherwise it is unstable.

\section{The case of competing nonlinearities: the results}\label{Sect:Bif-ComNon}

We consider Eq.\,(\ref{Eq:u_gen_pq}) for three types of nonlinearity:  $p=3$, $q=4$ (``$3:4$-model'', the cubic-quartic nonlinearity), $p=2$, $q=3$ (``$2:3$-model'', the quadratic-cubic nonlinearity) and $p=3$, $q=5$ (``$3:5$-model'', the cubic-quintic nonlinearity). Some information about these cases is given in  Table \ref{Tab:DataModels}.

\begin{table}
    \centering
    \begin{tabular}{c|c|c|c|c}
    \hline
        $p$,$q$ & $a$,$A$ & $\gamma^*$
        & $\gamma^{**}$ & $U(x)$\\[2mm]
        \hline
        $p=3$, $q=4$\quad &\begin{tabular}{c} positive roots of \\[-2mm] $\gamma u^3-u^2+1=0$\end{tabular}   & $\,\displaystyle \frac{5}{6\sqrt{6}}\,$  & \, $\displaystyle \frac{2}{3\sqrt{3}}$\, & no explicit form\\
        \hline
        $p=2$, $q=3$\quad &\,$\displaystyle \frac1{2\gamma}\left(1\pm\sqrt{1-4\gamma}\right)$\,       &  $\,\displaystyle \frac29$\, &\,  $\displaystyle \frac14$\, & \,$\displaystyle \frac{3}{1+\sqrt{1-9\gamma/2}\cosh x}$\,\\
        \hline
        $p=3$, $q=5$\quad &\,$\displaystyle \frac1{\sqrt{2\gamma}}\sqrt{1\pm\sqrt{1-4\gamma}}$\,    &\, $\displaystyle \frac3{16}$\, & \, $\displaystyle \frac14$ & $\displaystyle \frac{2}{\sqrt{1+\sqrt{1-16\gamma/3}\,\cosh 2x}}\,$\\
        \hline
    \end{tabular}
    \caption{Values of $a$, $A$ in anticontinuum limit, $\gamma^*$, $\gamma^{**}$ (see Sect. \ref{Sect:AL}) and the function $U(x)$, the continuous limit, (see Sect. \ref{Sect:ContLimit})}
    \label{Tab:DataModels}
\end{table}

We computed the branches of ILMs numerically using the method of ``pseudoarclength continuation'' \cite{Doedel07}. For our case, this method was improved for detecting points of pitchfork bifurcations and to find numerically the side branches of ILMs that are born at these points (see Appendix \ref{Sect:Numerics} for detail).
%It was found that all, except two, branches of ILMs undergo bifurcations and disappear at some value of $\alpha$.

\subsection{Tables of bifurcations}\label{Sec:tables}

Below we describe all the bifurcations of ILMs with codes of length 1,2 and 3. These results are presented in Tables \ref{Tab:a} -\ref{Tab:Rest}, see Appendix \ref{Sect:Tables}.  The following comments are in order.

(i) Quantitatively, Tables \ref{Tab:a} -\ref{Tab:Rest} cover {\it all the three cases}, i.e, $3:4$-model, $2:3$-model and $3:5$-model. The types of the bifurcations are the same for all the three equations. The difference lies in the critical values  $\gamma_k$, $k=0,1,\ldots$, that separate sub-intervals $\Delta_k$, $k=0,1,\ldots$,  with different behavior,  see the point (ii). 

(ii) Tables \ref{Tab:a} -\ref{Tab:Rest} show  switching between the branches of ILMs that occur when $\gamma$ varies. The interval $(0;\gamma^*)$ is divided into several sub-intervals, $\Delta_0=(0,\gamma_0)$, $\Delta_1=(\gamma_0,\gamma_1)$ etc. The bifurcations of branches are different at different $\Delta_k$, $k=0,1,\ldots$. The values of $\gamma_k$, $k=0,1,\ldots$, depend on the model and the branch of ILMs under consideration. For instance, (see Table \ref{Tab:a}, line 2), the branch of ILMs with code $(|A_+)$ being continued from ACL,  merges with the branch with code $(a_+|A_+a_+)$, if $\gamma\in \Delta_0$, and  with the branch with code $(|a_+)$, if $\gamma\in \Delta_1$. The bifurcation values of $\alpha$ when these branches merge also depend on  $p:q$-model and the code of ILMs. 

(iii) The branches $(a_+)$, $(a_+a_+)$ can be regarded as analogues of the  Sievers-Takeno mode and the Page mode of the DNLS equation. It follows from Table \ref{Tab:a}, lines 1 and 3, that they are $\infty$-branches at some sub-interval $\gamma\in \Delta_0$. This means that if $\gamma\in \Delta_0$ these branches do not bifurcate when $\alpha$ grows and when $\alpha\to\infty$ these ILMs approach the solution in continuum limit, see Sect.~\ref{Sect:ContLimit}.  The sub-intervals $\Delta_0$ are different for $(a_+)$ and $(a_+a_+)$ and depends on the model. Table \ref{Tab:a} includes all the bifurcations related to $(|a_+)$ and $(a_+|a_+)$ (see also Fig. \ref{Fig:Cascade}). It follows from Table \ref{Tab:a} (line 6) that the branch $(a_+|A_+a_+)$ is also an $\infty$-branch at some sub-interval of $\gamma$. Its relation with the branches $(|a_+)$ and $(a_+|a_+)$ is discussed in Sect. \ref{Sec:InftyBranches}.

\begin{figure}
    \centerline{\includegraphics[width=0.85\textwidth]{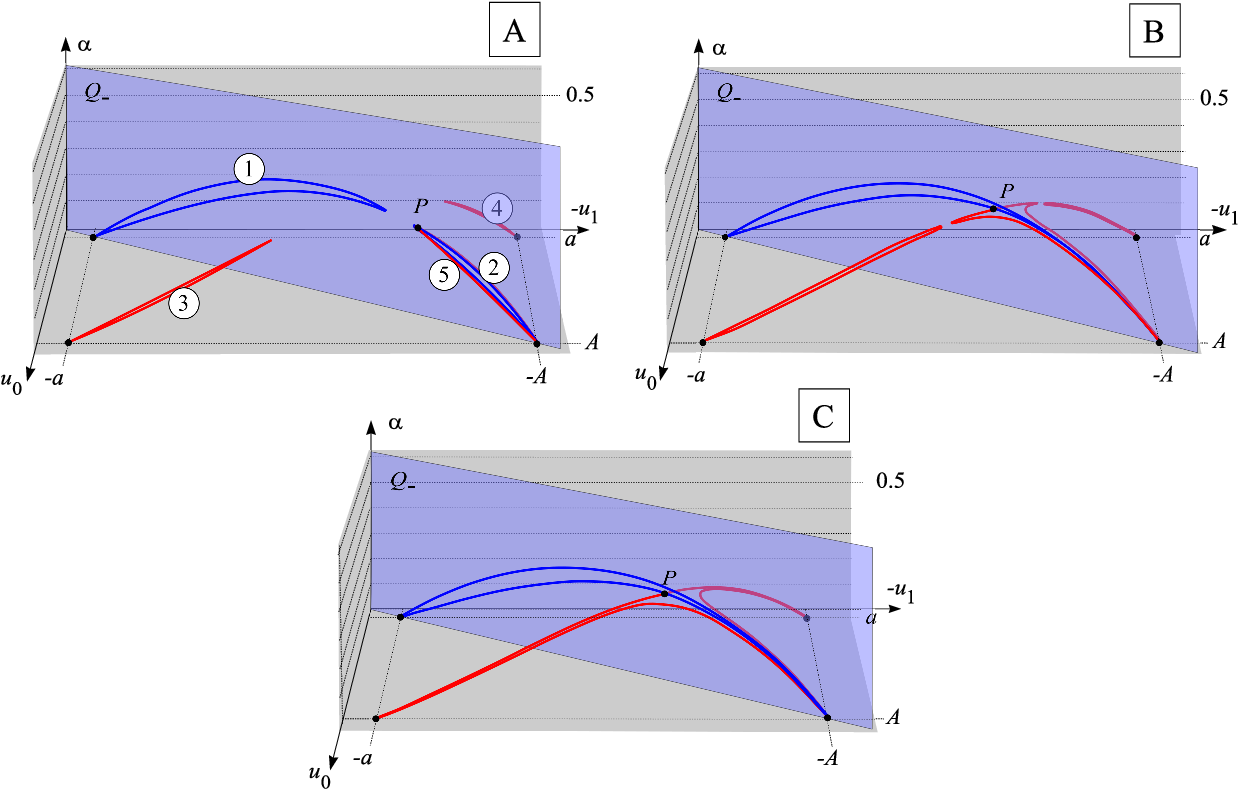}}
    \caption{Bifurcations of the ILMs,  $3:4$-model, see Table \ref{Tab:a+a-}. Panel (A), $\gamma\approx 0.2752$. $I_{\rm in}^-$-symmetric branches of ILMs $(a_+|a_-)$ and $(a_+a_+|a_-a_-)$ undergo the fold bifurcation and form a loop (blue, (1)).  $I_{\rm in}^-$-symmetric branches  $(A_+|A_-)$ and $(a_+A_+|A_-a_-)$ undergo the fold bifurcation and form a loop (blue, (2)). The nonsymmetric branches $(a_+|A_-)$ and $(a_+A_+|A_-)$ and their $I_{\rm in}^-$-symmetric counterparts $(A_+|a_-)$ and $(A_+|A_-a_-)$ merge, red loops, (3) and (4). The pair of branches $(a_+A_+|A_-)$ and $(A_+|A_-a_-)$ related by $I_{\rm in}^-$-symmetry  stick to the loop (2) (pitchfork bifurcation). Panel (B), $\gamma\approx0.2775$. The loops (1) and (2) reconnect: arise new $I_{\rm in}^-$-symmetric loops   $(a_+|a_-)\leftrightarrow(A_+|A_-)$ and $(a_+a_+|a_-a_-) \leftrightarrow (a_+A_+|A_-a_-)$. Panel (C), $\gamma\approx 0.2792$. The branch $(a_+|A_+a_-)$ reconnect with $(a_+|A_+A_-)$ and the branch $(a_+|A_-a_-)$ reconnects with $(A_+|A_-a_-)$ forming new loops. After this switching the branches $(a_+|A_-)$ and  $(A_+|a_-)$ stick to the loop $(a_+|a_-)\leftrightarrow(A_+|A_-)$ (pitchfork bifurcation).}
    \label{Fig:3d_real}
\end{figure}

(iv) Table \ref{Tab:a+a-}  shows the bifurcations related to modes $(a_+|a_-)$ and $(A_+|A_-)$. These modes are analogues of the first of the twisted modes for the DNLS equation, see the right lower panel of Fig. \ref{Fig:Basic_shapes}. This block of bifurcations is also illustrated by 3D view in Fig\ref{Fig:3d_real}.

(v)  Table \ref{Tab:a+0a-}  shows the bifurcations related to modes $(|a_+0a_-)$ and $(|A_+0A_-)$, that are analogues of the second of the twisted modes of the DNLS equation (see the right lower panel of Fig. \ref{Fig:Basic_shapes}).

(vi) Table \ref{Tab:Rest} shows the bifurcations of the rest of the ILMs that  have the length of the code 1,2 and 3 but were not included in Tables \ref{Tab:a}-\ref{Tab:a+0a-}.

(vii) Tables \ref{Tab:a} -\ref{Tab:Rest} include all the branches of ILMs with the codes of the length 1,2 and 3, up to the symmetries $I_{\rm in}^\pm$ and $I_{\rm on}^\pm$. The bifurcations of ILMs related to them by these symmetries can be also easily understood. For instance, merging of the branch $(|A_+)$ with the branch $(|a_+)$ (see line 2 of Table \ref{Tab:a}) implies merging of the branch $(|A_-)$ with the branch $(|a_-)$, etc.

\subsection{ILM without ACL counterpart}\label{Sect:without_ACL}

Unlike DNLS equation, new solutions of Eq.\,(\ref{Eq:u_gen_pq})  can bifurcate from the branches of ILMs. These ILMs have no ACL counterpart and consequently, they have no an ascribed code. For 3:5-model this fact was established in \cite{TD10} (see also \cite{ChP2011}). It was demonstrated that 3:5-model  admits so-called ``snaking'' phenomenon. This phenomenon implies the presence of families  of non-symmetric solutions that connect the branches of Sievers-Takeno and Page modes. Recently, the same phenomenon was found in $2:3$-model, \cite{Droplets02}. 

\begin{figure}
    \centerline{\includegraphics[width=0.9\textwidth]{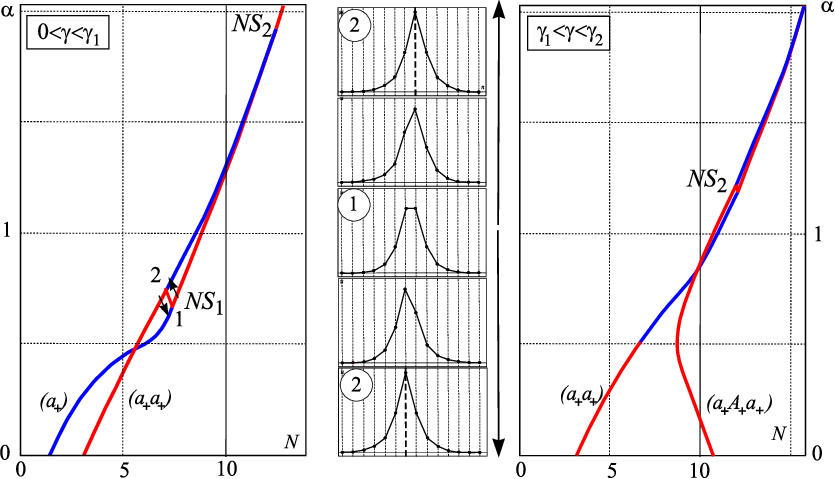}}
    \caption{Nonsymmetric connections between the branches $(|a_+)$ and $(a_+|a_+)$ (left panel) and $(a_+|a_+)$ and $(a_+|A_+a_+)$ (right panel), $N=\sum_k |u_k|$. Blue segments: stable ILMs, red segments: unstable ILMs. The connection $NS_1$ (marked $1\leftrightarrow 2$) is illustrated by central panel}
    \label{Fig:Asymmetric}
\end{figure}

We report on observation of similar families of nonsymmetric ILMs in 3:4-model also. This phenomenon is illustrated by Fig.~\ref{Fig:Asymmetric}.  Two families related by $I_{\rm in}^+$-symmetry exist for $0<\gamma<\gamma_{(a)}$, (where $\gamma_{(a)}$ is some critical value), see Fig.~\ref{Fig:Asymmetric}, left panel. They appear through pitchfork bifurcation from $(a_+|a_+)$ branch, an analog of the Page mode of DNLS equation. One of these families dies by sticking to $(|a_+)$ branch, another one - by sticking to $(a_+|)$ branch (note that ILMs on $(|a_+)$ and $(a_+|)$ are analogs of the Sievers-Takeno mode, one shifted with respect to another by one lattice spacing). There are many (presumably - infinitely many) such connections between the branches $(|a_+)$ and $(a_+|a_+)$. Two of them marked by  $NS_1$ and  $NS_2$ are shown in Fig.~\ref{Fig:Asymmetric}. 

For greater values of $\gamma$ the nonsymmetric connections between other  $\infty$-branches exist. For instance, for $\gamma\gtrapprox\gamma_{(a)}$ nonsymmetric connections exist between the $\infty$-branches $(a_+|a_+)$ and $(a_+|A_+a_+)$ (see right panel of Fig.~\ref{Fig:Asymmetric}). The transformation of there connections when $\gamma$ varies is discussed in in Sect. \ref{Sec:InftyBranches}.

\subsection{``Birth'' and ``death'' of $\infty$-branches when $\gamma$ varies}\label{Sec:InftyBranches}

The results of Sect.~\ref{Sec:tables} imply that ``most part'' of ILMs die  as $\alpha$ grows, so, they cannot be continued from ACM to continuum limit. 
For instance, the DNLS model admits only two $\infty$-branches, that are $(a_+)$ and $(a_+a_+)$. Here we present a scenario how $\infty$-branches appears and disappears passing through a cascade of bifurcations as $\gamma$ varies. This scenario is common for all the three models, i.e., $2:3$-model, $3:4$-model and $3:5$-model,
\begin{figure}
    \centerline{\includegraphics[width=0.9\textwidth]{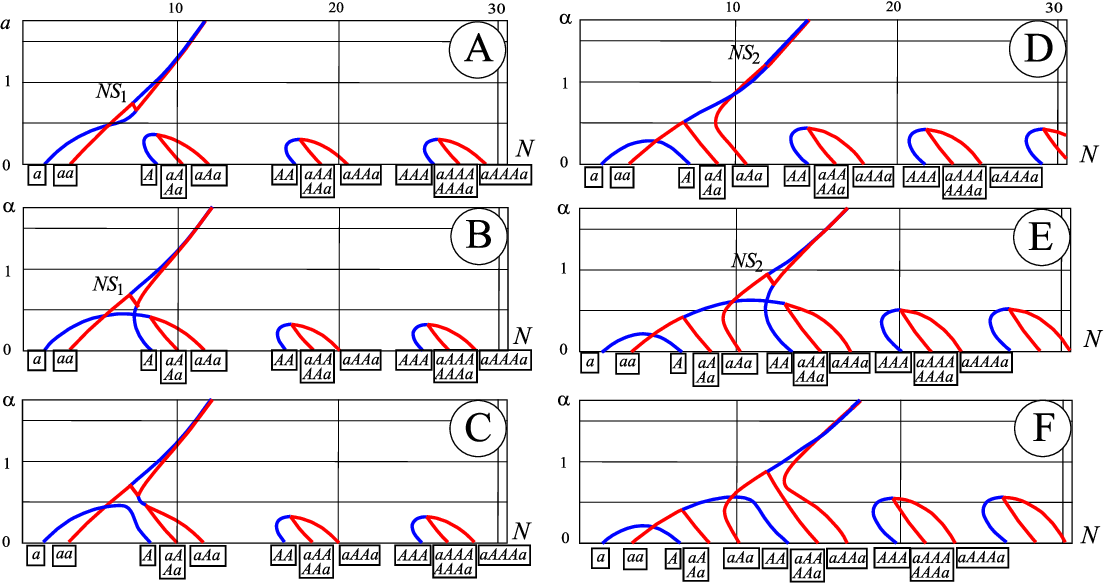}}
    \caption{The scenario how $\infty$-branches appear and disappear passing through a cascade of bifurcations as $\gamma$ varies, $N=\sum_k |u_k|$. Blue segments: stable ILMs, red segments: unstable ILMs. See comments in the text}
    \label{Fig:Cascade}
\end{figure}

Fig.~\ref{Fig:Cascade} gives a general idea of this scenario. 
\begin{itemize}
\item For $0<\gamma<\gamma_{(a)}$ the only $\infty$-branches are $(|a_+)$ and $(a_+|a_+)$ branches (see Fig.~\ref{Fig:Cascade} A). 
\item At $\gamma=\gamma_{(a)}$ the branch $(|a_+)$ 
touch the loop with ends $(|A_+)$ and $(a_+|A_+a_+)$. As a result, at $\gamma=\gamma_{(a)}$ the loop  with ends $(|a_+)$ and $(|A_+)$ arises and the branch $(a_+|A_+a_+)$ becomes  an $\infty$-branch.
Now the only $\infty$-branches are $(a_+|a_+)$ and $(a_+|A_+a_+)$.
\item At $\gamma=\gamma_{(Aa)}$ the branches $NS_1$ of nonsymmetric ILMs merge with $(|A_+)$ and the loop  with ends $(|a_+)$ and $(|A_+)$ appears (see Fig.~\ref{Fig:Cascade} B,C). Then pair of branches $NS_1$ of nonsymmetric solutions merge with the pair  $(a_+|A_+)$, $(A_+|a_+)$ (see Fig.~\ref{Fig:Cascade} C,D). 
\item  At $\gamma=\gamma_{(aa)}$ the branch $(a_+|a_+)$ touches the loop with ends $(A_+|A_+)$ and $(a_+A_+|A_+a_+)$. After switching the loop with ends $(a_+|a_+)$ and $(A_+|A_+)$ forms. The branch $(a_+A_+|A_+a_+)$ becomes  an $\infty$-branch. The only $\infty$-branches now are $(a_+|A_+a_+)$ and $(a_+A_+|A_+a_+)$.
\item At $\gamma=\gamma_{(AAa)}$ the branches $NS_2$  merge with $(a_+A_+|A_+)$ and $(A_+|A_+a_+)$ (see Fig.~\ref{Fig:Cascade} E,F). 
\item The procedure continues. Pairs of $\infty$-branches arise as follows
\small
\begin{tabular}{cccccccc}
  &  $\gamma_{(a)}$ & & $\gamma_{(aa)}$ &&  $\gamma_{(aAa)}$ && $\ldots$\\
$\left[\begin{tabular}{c}
$(|a_+)$\\
$(a_+|a_+)$
\end{tabular}
\right]$
&$\longrightarrow$ & 
$\left[\begin{tabular}{c}
$(a_+|A_+a_+)$\\
$(a_+|a_+)$
\end{tabular}
\right]$
&$\longrightarrow$ & 
$\left[\begin{tabular}{c}
$(a_+|A_+a_+)$\\
$(a_+A_+|A_+a_+)$
\end{tabular}
\right]$
&$\longrightarrow$ & 
$\left[
\begin{tabular}{c}
$(a_+A_+|A_+A_+a_+)$\\
$(a_+A_+|A_+a_+)$
\end{tabular}
\right]$
&$\longrightarrow$
\end{tabular}
\normalsize
\end{itemize}

\subsection{Stability}\label{Sect:stability}

Unfortunately, no decisive rule is known to conclude whether an ILM is stable or not. It follows from Fig.\ref{Fig:Cascade} that the stability of even the simplest ILMs (like the Sievers-Takeno mode and the Page mode) can vary when $\alpha$ grows. Below we report on few observations on stability/instability of ILMs only for the modes that are close to ACL (i.e., for  $\alpha\in(0;\tilde\alpha)$ where $\tilde\alpha$ is small enough).
The results below are valid for all the three models, i.e., $2:3$-, $3:5$- and $3:5$-models. Evidently, the ILMs related to ones presented below by symmetries $I_{\rm in}^\pm$ and $I_{\rm on}^\pm$ keep also stability/unstability.
\begin{itemize}
    \item Among the ILMs of the codes of length 1 both the ILMs $(a_+)$ and $(A_+)$ are stable.
    \item Among the ILMs of the codes of length 2 the ILMs $(a_+a_-)$, $(A_+,a_-)$ and $(A_+A_+)$ are stable.
    \item Among the ILMs of the codes of length 3 the ILMs $(A_+A_+A_+)$, $(A_+0A_+)$, $(A_+A_+a_-)$, $(a_+0A_-)$, $(a_+0a_-)$ and $(a_+a_-a_+)$ are stable. The ILM $(a_+a_+A_-)$ can exhibit stability at some values of $\gamma$ and instability at others.
    \item Presumably, the modes with codes of alternating symbols $a_\pm$ and non-alternating symbols $A_\pm$ (for instance, $(a_+a_-a_+a_-)$ and $(A_+A_+A_+A_+)$) are stable). Contrarily, the modes with codes of non-alternating symbols $a_\pm$ and alternating symbols $A_\pm$  are unstable.
\end{itemize}
More detailed analysis of stability for ILMs close to  ACL will be presented in separate publication.

\section{Conclusion}\label{Sect:Concl}

In the paper, we consider the problem of possible types of nonlinear localized modes (ILMs) that can be described by DNLS-type equation with competing nonlinearities. The models of such kind are of considerable interest for the Bose-Einstein Condensate theory. Three cases are considered, called $3:4$-model (i.e., cubic-quintic model), $2:3$-model (i.e., quadratic-cubic model) and $3:5$-model (i.e., cubic-quintic model). These models differ by the degrees of competing terms in the nonlinearity. All of them differ from classical DNLS model by presence of {\it five} equilibria (instead of three ones in the DNLS case).  In our study, the main emphasis is on the 3:4-model that has been the less studied between these three ones. 

A peculiarity of the situation is the presence of additional parameter $\gamma$ that describe the balance between the competing powers. Our approach employs numerical continuation of ILMs from the anti-continuum limit where the coupling between the lattice sites is neglected  ($\alpha=0$).  We analyze the bifurcations for the simplest mode configurations for all the three models and their dependence on $\gamma$. Our study shows that the ``most part'' of branches originated at the anti-continuum limit bifurcate and do not exist for large values of $\alpha$.  We present comprehensive tables of bifurcations for the ILMs that involve not more than 3 excited lattice sites. Qualitatively, these tables are identical for all the three models.

It has been known that the DNLS with competing nonlinearity may admit branches of ILMs that do not exist at the anti-continuum limit. Before, these solutions have been found for quadratic-cubic and for cubic-quintic models. We show that the same phenomenon takes place also in cubic-quartic model. This fact is promising for existence of moving ILMs in cubic-quartic model when $\alpha$ takes some peculiar values (so-called ``transparence points'', see \cite{MCKC06,MCKC08,AKLP19,AT19}).    

Also we analyze in detail the  branches of ILMs that admit the continuation from anti-continuum limit ($\alpha=0$) to continuum limit ($\alpha\to\infty$). We called these branches $\infty$-branches. Only two $\infty$-branches exist in the DNLS case, they are the branches of the Sievers-Takeno modes and the Page modes. In the case of competing nonlinearity the situation is more complex. At any value of $\gamma$ also only two $\infty$-branches exist. However, when $\gamma$ grows sequential switching of $\infty$-branches occurs: one $\infty$-branch merge with another branch forming a loop, but similtaneously new $\infty$-branch arises. As a result,  ILMs at $\infty$-branch involve more and more excited sites when $\gamma$ grows. 

Finally, we study the stability of ILMs localized at one, two and three lattice sites and established the modes that are stable in all the three models. 

Since our study is basically numerical, it would be quite desirable to support the findings by mathematical statements. In fact, we found that the three models under consideration have quite similar properties. For instance, we conjecture that the lists of bifurcations given in Tables \ref{Tab:a}-\ref{Tab:Rest} (see Appendix \ref{Sect:Tables}) are quite generic for wide class of competing nonlinearities. Another interesting issue for the further study is the stability of ILMs. A challenging problem for this study is to establish a relation between the code of the ILM and  its stability/instability, at least for small values of $\alpha$, like it was done in \cite{PKF05}.

\section*{ACKNOWLEDGMENTS}

GLA and PAK are grateful to the team of the Laboratory of Theoretical Physics of
Physical-Technical Institute of the Uzbekistan Academy of Sciences for hospitality and discussion, with special thanks to Prof. Eduard N. Tsoy.

\appendix

\section{Numerical algorithm}\label{Sect:Numerics}

For computation of ILMs we employed the method of {\it pseudoarclength continuation} \cite{Doedel07}. Being launched from ACL ($\alpha=0$), it allows to compute branches of ILMs for $\alpha>0$ and to pass the points of the fold bifurcation. This provides a way to detect loops made up of two joined branches of ILMs. Besides this, we need the option to fix the points of secondary (pitchfork) bifurcations and to compute the branches that arise at these points. Briefly, the algorithm that we use can be described as follows. 

Let ${\bf F}: \mathbb{R}^N\times \mathbb{R} \to \mathbb{R}^N$, $\lambda\in\mathbb{R}$,  and one seeks for $\alpha$-depended branch of solutions of vector equation 
\begin{gather}
{\bf F}({\bf x},\lambda)=0.
\label{BasicEq}
\end{gather}
Let $\Gamma$ be a curve in $\mathbb{R}^N\times \mathbb{R}$ that correspond to this branch and $U_0=({\bf x}_0,\lambda_0)\in\Gamma$. Introduce
\begin{gather*}
{\bf F}'_{\bf x}({\bf x}_0,\lambda_0)=\left(
\begin{array}{cccc}
F'_{1,x_1}({\bf x}_0,\lambda_0)&F'_{1,x_2}({\bf x}_0,\lambda_0)&\ldots&F'_{1,x_N}({\bf x}_0,\lambda_0)\\[2mm]
F'_{2,x_1}({\bf x}_0,\lambda_0)&F'_{2,x_2}({\bf x}_0,\lambda_0)&\ldots&F'_{2,x_N}({\bf x}_0,\lambda_0)\\[2mm]
\vdots &                  \vdots                   &\ddots& \vdots\\[2mm]
F'_{N,x_1}({\bf x}_0,\lambda_0)&F'_{N,x_2}({\bf x}_0,\lambda_0)&\ldots&F'_{N,x_N}({\bf x}_0,\lambda_0)
\end{array}
\right),
\end{gather*}
and
\begin{gather*}
{\bf F}'_{\lambda}({\bf x}_0,\lambda_0)=\left(
\begin{array}{c}
F'_{1,\lambda}({\bf x}_0,\lambda_0)\\[2mm]
F'_{2,\lambda}({\bf x}_0,\lambda_0)\\[2mm]
\vdots\\[2mm]
F'_{N,\lambda}({\bf x}_0,\lambda_0)
\end{array}
\right).
\end{gather*}
To shorten the notation we denote $F'_{n,x_m}$ the derivative of the component $F_n$ with respect to $x_m$ and $F'_{n,\lambda}$ the corresponding derivative with respect to $\lambda$.

1. Let the rank of  augmented $(N+1)\times N$ matrix
\begin{gather*}
{\bf F}'({\bf x}_0,\lambda_0)=\left(
\begin{array}{c|c}
{\bf F}'_{\bf x}({\bf x}_0,\lambda_0)& {\bf F}'_\lambda({\bf x}_0,\lambda_0)
\end{array}
\right)
\end{gather*}
is equal to $N$. Then the unit tangent vector $T_0=({\bf X}_0,\Lambda_0)$ to $\Gamma$ taken in $U_0$ satisfies the relations 
\begin{align}
&{\bf F}'_{\bf x}({\bf x}_0,\lambda_0){\bf X}_0+ {\bf F}'_{\lambda}({\bf x}_0,\lambda_0) \Lambda_0=0,\label{Tangent}\\[2mm]
&\|{\bf X}_0\|^2+\Lambda^2=1,\nonumber
\end{align}
and can be computed unambiguously up to sign. We seek for the next point $U_1=({\bf x}_1,\lambda_1)\in\Gamma$ on the hyperplane $P$ that is orthogonal to $T_0$ and is situated on the distance $\Delta s$ from $U_0$. $\Delta s$ can be regarded as a step along the curve $\Gamma$. Then $U_1$ satisfies (\ref{BasicEq}) and the equation
\begin{gather}
  {\bf X}_0^T({\bf x}_1-{\bf x}_0)+\Lambda_0(\lambda_1-\lambda_0)-\Delta s=0. \label{InPlane} 
\end{gather}
We solve the system (\ref{BasicEq}), (\ref{InPlane}) by means of the Newton method, starting with the initial guess
\begin{gather*}
{\bf x}_1^{(0)}={\bf x}_0+\Delta s\cdot {\bf X}_0,\quad \lambda_1^{(0)}=\lambda_0+\Delta s\cdot \Lambda_0.
\end{gather*}
Then the procedure continues with $U_1$ instead of $U_0$.

2. If the point $U_0$ is a point of pitchfork bifurcation then the rank of  the matrix ${\bf F}'_{\bf x}({\bf x}_0,\lambda_0)$ is equal to $N-1$ and the rank of  $(N+1)\times(N+1)$ matrix,
\begin{align*}
{\cal F}({\bf x}_0,\lambda_0)=\left
(\begin{array}{cccc}
{\bf F}'_{\bf x}({\bf x}_0,\lambda_0)&\quad{\bf F}'_{\lambda}({\bf x}_0,\lambda_0)\\[2mm]
{\bf X}_0&\Lambda_0
\end{array}
\right)
\end{align*} 
is $N$. Let ${ Z}_0=(\tilde{\bf X}_0,\tilde{\Lambda}_0)$ be the normalized eigenvector that corresponds to zero eigenvalue of ${\cal F}({\bf x}_0,\lambda_0)$. It is straightforward to check that ${Z}_0$ is orthogonal to $T_0$. In order to find side branches $\tilde\Gamma_{1,2}$ that arise due to pitchfork bifurcation we repeat the procedure described in p.1 but taking vector $Z_0$ instead of $T_0$. Switching to the branch $\tilde\Gamma_{1}$ or $\tilde\Gamma_{2}$ depends on the choice of $Z_0$ or $-Z_0$. For a point $\tilde {U}_1=(\tilde{\bf x}_1,\tilde\lambda_1)$ situated on a curve $\tilde\Gamma_1$  we have ${\bf F}(\tilde{\bf x}_1,\tilde\lambda_1)=0$ and 
\begin{gather}
\tilde{\bf X}^T_0({\bf \tilde x}_1-{\bf x}_0)+\tilde{\Lambda}_0(\tilde{\lambda}_1-\lambda_0)-\Delta s=0,
\end{gather}
where $\Delta s$ is a step along the curve $\tilde\Gamma_1$.

In practice, in vicinity of the point of pitchfork bifurcation the matrix ${\cal F}$ is ``nearly'' degenerate, so, there is a problem of convergence of the Newton algorithm. When moving along the main curve $\Gamma$, we passed this point by ``jumping over''  taking the step $\Delta s$ large  enough.  The switching to  the secondary branches implies the following trick:
\begin{itemize}
\item When following the main curve $\Gamma$, compute determinant of ${\cal F}$  at each point. Fix the situation when ${\rm det} {\cal F}$ changes sign and the point $({\bf x}_0,\lambda_0)$ where $\left|{\rm det} {\cal F}\right|$ takes the smallest value;
\item Compute the eigenvector of ${\cal F}$  at $({\bf x}_0,\lambda_0)$ that corresponds to the eigenvalue that has the smallest absolute value;
\item use this eigenvector instead of $Z_0$ in the procedure of seeking of  $\tilde{\Gamma}_{1,2}$.    
\end{itemize}

\newpage

\section{Tables of bifurcations for ILMs (the codes of length $\le 3$)}\label{Sect:Tables}

\begin{table}[h]
\begin{tabular}{c|c|c}
\hline
&Code & Counterparts ($\gamma\in\Delta_0$, $\gamma\in \Delta_1,\ldots$)\\
\hline
\hline
1\phantom{10} &
\begin{tabular}{l}
{$(|a_+)$}
\end{tabular}
 &
\begin{tabular}{p{60mm}|p{60mm}}
\quad $\infty$& \quad$(|A_+)$\quad (f.) 
\end{tabular}
\\
\hline
%---------------------------------------------------
2\phantom{10}&
\begin{tabular}{l}
{$(|A_+)$}
\end{tabular}
&
\begin{tabular}{p{60mm}|p{60mm}}
\quad {$(a_+|A_+a_+)$}\quad  (f.)&
\quad$(|a_+)$\quad (f.) 
\end{tabular}
\\
\hline
%---------------------------------------------------
3\phantom{10}&
\begin{tabular}{l}
{$(a_+|a_+)$}
\end{tabular}&
\begin{tabular}{p{60mm}|p{60mm}}
\quad $\infty$ & \quad$(A_+|A_+)$\quad (f.) 
\end{tabular}
\\
%---------------------------------------------------
\hline
4\phantom{10}&
\begin{tabular}{l}
{$(A_+|A_+)$}
\end{tabular}
&
\begin{tabular}{p{60mm}|p{60mm}}
\quad {$(a_+A_+|A_+a_+)$}\quad  (f.)\quad& \quad$(a_+|a_+)$\quad (f.) 
\end{tabular}
\\
\hline
5\phantom{10} &\begin{tabular}{l} {$(a_+|A_+)$},\\[-3mm]
 {$(|A_+a_+)$}
\end{tabular}
 &
\begin{tabular}{p{29mm}|p{29mm}|p{60mm}}
$\left(
\begin{tabular}{l}
{$(a_+|A_+a_+)$}\\[-2mm]
{$(|A_+)$}
\end{tabular}
\right.$(p.) & 
$\left(
\begin{tabular}{l}
{$(a_+|A_+a_+)$}\\[-2mm]
$\infty$
\end{tabular}
\right.$(p.) & 
\end{tabular}
\\
%---------------------------------------------------
\cline{2-3}
&\begin{tabular}{l} {$(a_+|A_+)$},\\[-3mm]
{$(A_+|a_+)$}
\end{tabular}
 &
\begin{tabular}{p{61mm}|p{30mm}|p{30mm}}
 &
$\left(
\begin{tabular}{l}
{$(a_+|a_+)$}\\[-2mm]
$\infty$
\end{tabular}
\right.$(p.) & 
$\left(
\begin{tabular}{l}
{$(a_+|a_+)$}\\[-2mm]
{$(A_+|A_+)$}
\end{tabular}
\right.$(p.) \\
\end{tabular}\\
\hline
%---------------------------------------------------
6\phantom{10}&
\begin{tabular}{l}
{$(a_+|A_+a_+)$}
\end{tabular}
 &
\begin{tabular}{p{40mm}|p{40mm}|p{40mm}}
\quad$(|A_+)$\quad (f.) &
\quad $\infty$\quad &
\quad$(A_+|A_+A_+)$\quad (f.)\\
\end{tabular}
\\
%----------------------------------------------
\hline
7\phantom{10}&
\begin{tabular}{l}
{$(A_+|A_+A_+)$}
\end{tabular}
 &
\begin{tabular}{p{60mm}|p{60mm}}
\quad$(a_+A_+|A_+A_+a_+)$\quad (f.) & \quad$(a_+|A_+a_+)$\quad (f.) 
\end{tabular}
\\
%----------------------------------------------
\hline

8\phantom{10} &\begin{tabular}{l} {$(A_+|A_+a_+)$},\\[-3mm]
{$(a_+A_+|A_+)$}
\end{tabular}
 &

\begin{tabular}{p{30mm}|p{29mm}|p{60mm}}
\fontsize{9}{12}\selectfont
$\left(
\begin{tabular}{l}
{$(a_+A_+|A_+a_+)$}\\[-2mm]
{$(A_+|A_+)$}
\end{tabular}
\right.$(p.)& 
\fontsize{9}{12}\selectfont
$\left(
\begin{tabular}{l}
{$(a_+A_+|A_+a_+)$}\\[-2mm]
$\infty$
\end{tabular}
\right.$(p.)& 
\end{tabular}\\
%---------------------------------------------------
\cline{2-3}
&\begin{tabular}{l} {$(A_+|A_+a_+)$},\\[-3mm]
{$(a_+|A_+A_+)$}
\end{tabular}
 &
\begin{tabular}{p{62mm}|p{30mm}|p{30mm}}
 &
$\left(
\begin{tabular}{l}
{$(a_+|A_+a_+)$}\\[-2mm]
$\infty$
\end{tabular}
\right.$(p.)& 
$\left(
\begin{tabular}{l}
{$(a_+|A_+a_+)$}\\[-2mm]
{$(A_+|A_+A_+)$}
\end{tabular}
\right.$(p.)\\
\end{tabular}\\
\hline
%---------------------------------------------------

\hline

\end{tabular}
\caption{The bifurcations of the branches of ILMs related to the branches $(|a_+)$ and $(a_+|a_+)$. Line 1: the branch with code $(|a_+)$ is an $\infty$-branch if $\gamma\in\Delta_0$. It merges at some finite $\alpha$ with branch $(|A_+)$ if $\gamma\in\Delta_1$. The mark (f.) means the fold bifurcation. Lines 2-4,6,7 are interpreted in the same way. Line 5: the branch of ILMs with code $(a_+|A_+)$ together with its $I_{\rm on}^+$-counterpart $(|A_+a_+)$ coalesces into the $I_{\rm on}^+$-symmetric loop with endpoints $(a_+|A_+a_+)$ and $(|A_+)$ through a pitchfork bifurcation (marked (p.) in the table). At some critical value of $\gamma$ this loop transforms into $I_{\rm on}^+$-symmetric $\infty$-branch $(a_+|A_+a_+)$. For greater values of $\gamma$, the branch  $(a_+|A_+)$ together with its $I_{\rm in}^+$-counterpart $(A_+|a_+)$ coalesces into $I_{\rm in}^+$-symmetric $\infty$-branch $(a_+|a_+)$ through a pitchfork bifurcation. For even larger values of $\gamma$ this $\infty$-branch transforms into  $I_{\rm on}^+$-symmetric loop with endpoints $(a_+|A_+a_+)$ and $(A_+|A_+A_+)$ and the pair $(a_+|A_+)$ and $(A_+|a_+)$ coalesce into it through a pitchfork bifurcation. Line 8 is interpreted in the similar way. For more information see Sect. \ref{Sec:InftyBranches}, Fig. \ref{Fig:Cascade}.}
\label{Tab:a}
\end{table}

\begin{table}
\begin{tabular}{c|c|c}
\hline
&Code & Counterparts ($\gamma\in\Delta_0$, $\gamma\in \Delta_1,\ldots$)\\
\hline
\hline
1\phantom{10} &
\begin{tabular}{l}
{$(a_+|a_-)$}
\end{tabular}
 &
\begin{tabular}{p{60mm}|p{60mm}}
\quad $(a_+a_+|a_-a_-)$\quad(f.)&\quad $(A_+|A_-)$\quad(f.)\quad 
\end{tabular}
\\
\hline
%---------------------------------------------------
2\phantom{10}&
\begin{tabular}{l}
{$(A_+|A_-)$}
\end{tabular}
&
\begin{tabular}{p{60mm}|p{60mm}}
\quad {$(a_+A_+|A_-a_-)$}\quad  (f.)\quad  &\quad {$(a_+|a_-)$}\quad  (f.)
\end{tabular}
\\
\hline
%---------------------------------------------------
3\phantom{10}&\begin{tabular}{l} {$(a_+|A_-)$},\\[-3mm]
{$(A_+|a_-)$}
\end{tabular}
 &
\begin{tabular}{p{60mm}|p{60mm}}
\quad\begin{tabular}{l}$(a_+|A_-a_-)$\quad (f.)\\[-2mm]
$(a_+A_+|a_-)$ \quad (f.) 
\end{tabular}& 
$\left(
\begin{tabular}{l}
{$(a_+|a_-)$}\\[-2mm]
{$(A_+|A_-)$}
\end{tabular}
\right.$\quad (p.) 
\end{tabular}
\\
\hline
%---------------------------------------------------
4\phantom{10}
&\begin{tabular}{l} {$(a_+a_+|a_-)$},\\[-3mm]
{$(a_+|a_-a_-)$}
\end{tabular}
&
\begin{tabular}{p{40mm}|p{40mm}|p{40mm}}
$\left(
\begin{tabular}{l}
{$(a_+|a_-)$}\\[-2mm]
{$(a_+a_+|a_-a_-)$}
\end{tabular}
\right.$\quad (p.) & 
$\left(
\begin{tabular}{l}
{$(a_+a_+|a_-a_-)$}\\[-2mm]
{$(a_+A_+|A_-a_-)$}
\end{tabular}
\right.$\quad (p.) & 
\begin{tabular}{l} $(a_+a_+|A_-)$\quad (f.)\\[-2mm]
$(A_+|a_-a_-)$ \quad (f.) 
\end{tabular}
\end{tabular}
\\
\hline
%---------------------------------------------------
5\phantom{10}&\begin{tabular}{l}{$(a_+|A_+a_-)$}
\end{tabular}
 &
\begin{tabular}{p{60mm}|p{60mm}}
\quad\begin{tabular}{l} $(|A_+a_-)$ \quad (f.) 
\end{tabular}& 
\quad\begin{tabular}{l} $(a_+|A_+A_-)$ \quad (f.) 
\end{tabular}
\end{tabular}
\\
\hline
%---------------------------------------------------
6\phantom{10}&\begin{tabular}{l} {$(|a_+a_+A_-)$}
\end{tabular}
 &
\begin{tabular}{p{60mm}|p{60mm}}
\quad\begin{tabular}{l} $(|a_+a_+A_-a_-)$\quad (f.) 
\end{tabular}& 
\quad\begin{tabular}{l} $(|a_+a_+a_-)$\quad (f.)\end{tabular}
\end{tabular}
\\
\hline
%----------------------------------------------
7\phantom{10}&\begin{tabular}{l} {$(a_+A_+|A_-)$},\\[-3mm]
{$(A_+|A_-a_-)$}
\end{tabular}
&
\begin{tabular}{p{40mm}|p{40mm}|p{40mm}}
$\left(
\begin{tabular}{l}
{$(A_+|A_-)$}\\[-2mm]
{$(a_+A_+|A_-a_-)$}
\end{tabular}
\right.$\quad (p.) &
$\left(
\begin{tabular}{l}
{$(a_+|a_-)$}\\[-2mm]
{$(A_+|A_-)$}
\end{tabular}
\right.$\quad (p.) & 
\begin{tabular}{l} $(a_+A_+|a_-)$\quad (f.)\\[-2mm]
$(a_+|A_-a_-)$ \quad (f.) 
\end{tabular}
\end{tabular}
\\
\hline
%----------------------------------------------

\hline

\end{tabular}
\caption{Bifurcations of branches of ILMs related to branch $(a_+|a_-)$ (an analog of the first twisted mode of the DNLS). Comments to lines 1,2,5 and 6  are similar to ones for lines 1-4 of Table \ref{Tab:a}. Line 3: for $\gamma$ below some threshold, $I_{\rm in}^-$-symmetric branches $(a_+|A_-)$ and $(A_+|a_-)$ merge with $I_{\rm in}^-$-symmetric branches $(a_+|A_-a_-)$ and $(a_+A_+|a_-)$ respectively. Above this critical value, these branches merge into a $I_{\rm in}^-$-symmetric loop formed by branches $(a_+|a_-)$ and $(A_+|A_-)$ (a pitchfork bifurcation). Lines 4 and 7 are interpreted in a similar way. See a visualization in Fig. \ref{Fig:3d_real}. }
\label{Tab:a+a-}
\end{table}

\begin{table}
\begin{tabular}{c|c|c}
\hline
&Code & Counterparts ($\gamma\in\Delta_0$, $\gamma\in \Delta_1,\ldots$)\\
\hline
\hline
%----------------------------------------------
1\phantom{10} &
\begin{tabular}{l}
{$(a_+|0a_-)$}
\end{tabular}
 &
\begin{tabular}{p{60mm}|p{60mm}}
\quad $(a_+a_+|0a_-a_-)$\quad(f.)&\quad $(A_+|0A_-)$\quad(f.)\quad 
\end{tabular}
\\
\hline
%---------------------------------------------------
2\phantom{10}&
\begin{tabular}{l}
{$(A_+|0A_-)$}
\end{tabular}
&
\begin{tabular}{p{60mm}|p{60mm}}
\quad {$(a_+A_+|0A_-a_-)$}\quad  (f.)\quad  &\quad {$(a_+|0a_-)$}\quad  (f.)
\end{tabular}
\\
\hline
%---------------------------------------------------
3\phantom{10}&\begin{tabular}{l} {$(a_+|0A_-)$},\\[-3mm]
{$(A_+|0a_-)$}
\end{tabular}
 &
\begin{tabular}{p{60mm}|p{60mm}}
\begin{tabular}{l} \quad $(a_+|0A_-a_-)$\quad (f.)\\[-2mm]
\quad $(a_+A_+|0a_-)$\quad (f.) 
\end{tabular}&
$\left(
\begin{tabular}{l}
{$(a_+|0a_-)$}\\[-2mm]
{$(A_+|0A_-)$}
\end{tabular}
\right.$\quad (p.) 
\end{tabular}
\\
\hline
\end{tabular}
\caption{Bifurcations of branches of ILMs related to branch $(a_+|0a_-)$ (an analog of the second twisted mode of the DNLS). The notation is the same as in Tables \ref{Tab:a+a-} and \ref{Tab:a+0a-}. }
\label{Tab:a+0a-}
\end{table}

\begin{table}
\begin{tabular}{c|c|c}
\hline
&Code & Counterparts ($\gamma\in\Delta_0$, $\gamma\in \Delta_1,\ldots$)\\
\hline
\hline
1\phantom{10} &
\begin{tabular}{l}
{$(a_+|0a_+)$}
\end{tabular}
 &
\begin{tabular}{p{60mm}|p{60mm}}
\quad $(a_+|a_+a_+)$\quad(f.)&\quad $(A_+|0A_+)$\quad(f.)\quad 
\end{tabular}
\\
\hline
%---------------------------------------------------
2\phantom{10}&
\begin{tabular}{l}
{$(a_+|a_+a_+)$}
\end{tabular}
&
\begin{tabular}{p{60mm}|p{60mm}}
\quad {$(a_+|0a_+)$}\quad  (f.)\quad  &\quad {$(A_+|a_+A_+)$}\quad  (f.)
\end{tabular}
\\
\hline
%---------------------------------------------------
3\phantom{10}&\begin{tabular}{l} {$(a_+|0A_+)$},\\[-3mm]
{$(A_+|0a_+)$}
\end{tabular}
 &
\begin{tabular}{p{40mm}|p{40mm}|p{40mm}}
\begin{tabular}{l} \quad $(a_+|a_+A_+)$\quad (f.)\\[-2mm]
\quad $(A_+|a_+a_+)$ \quad (f.) 
\end{tabular}& 
$\left(
\begin{tabular}{l}
{$(a_+|0a_+)$}\\[-2mm]
{$(a_+|a_+a_+)$}
\end{tabular}
\right.$\quad (p.) 
&
$\left(
\begin{tabular}{l}
{$(a_+|0a_+)$}\\[-2mm]
{$(A_+|0A_+)$}
\end{tabular}
\right.$\quad (p.) 
\end{tabular}
\\
\hline
%---------------------------------------------------
4\phantom{10}
&\begin{tabular}{l} {$(a_+|a_+A_+)$},\\[-3mm]
{$(A_+|a_+a_+)$}
\end{tabular}
&
\begin{tabular}{p{40mm}|p{40mm}|p{40mm}}
\begin{tabular}{l} \quad $(a_+|0A_+)$\quad (f.)\\[-2mm]
\quad $(A_+|0a_+)$ \quad (f.) 
\end{tabular}& 
$\left(
\begin{tabular}{l}
{$(a_+|0a_+)$}\\[-2mm]
{$(a_+|a_+a_+)$}
\end{tabular}
\right.$\quad (p.) 
&
$\left(
\begin{tabular}{l}
{$(a_+|a_+a_+)$}\\[-2mm]
{$(A_+|a_+A_+)$}
\end{tabular}
\right.$\quad (p.) 
\end{tabular}
\\
\hline
%---------------------------------------------------
5\phantom{10}&\begin{tabular}{l}
{$(A_+|0A_+)$}
\end{tabular}
 &
\begin{tabular}{p{60mm}|p{60mm}}
\quad $(A_+|a_+A_+)$\quad(f.)&\quad $(a_+|0a_+)$\quad(f.)\quad 
\end{tabular}
\\
\hline
%---------------------------------------------------
6\phantom{10}&\begin{tabular}{l}
{$(A_+|a_+A_+)$}
\end{tabular}
 &
\begin{tabular}{p{60mm}|p{60mm}}
\quad $(A_+|0A_+)$\quad(f.)&\quad $(a_+|a_+a_+)$\quad(f.)\quad 
\end{tabular}
\\
\hline
\hline

%----------------------------------------------
1\phantom{10}&
 \begin{tabular}{l}
{$(a_+|a_-a_+)$}
\end{tabular}
 &
\begin{tabular}{p{60mm}|p{60mm}}
\quad$(a_+a_+|a_-a_+a_+)$\quad (f.) & \quad ${(a_+|A_-a_+)}$\quad (f.)
\end{tabular}
\\
\hline
%----------------------------------------------
2\phantom{10}&
\begin{tabular}{l}
{$(a_+|A_-a_+)$}
\end{tabular}
 &
\begin{tabular}{p{60mm}|p{60mm}}
\quad $(a_+a_+|A_-a_+a_+)$\quad (f.)  & \quad$(a_+|a_-a_+)$\quad (f.) 
\end{tabular}
\\
\hline
\hline
%----------------------------------------------
1\phantom{10}&
 \begin{tabular}{l}
{$(|A_+a_+a_-)$}
\end{tabular}
 &
\begin{tabular}{p{60mm}|p{60mm}}
\quad$(a_+|A_+a_+a_-)$\quad (f.) & \quad ${(|A_+a_+A_-)}$\quad (f.)
\end{tabular}
\\
\hline
%----------------------------------------------
2\phantom{10}&
\begin{tabular}{l}
{$(|A_+a_+A_-)$}
\end{tabular}
 &
\begin{tabular}{p{60mm}|p{60mm}}
\quad $(a_+|A_+a_+A_-)$\quad (f.)  & \quad$(|A_+a_+a_-)$\quad(f.) 
\end{tabular}
\\
\hline 
\hline
%----------------------------------------------
1\phantom{10}&
 \begin{tabular}{l}
{$(|A_+a_-a_+)$}
\end{tabular}
 &
\begin{tabular}{p{60mm}|p{60mm}}
\quad$(a_+|A_+a_-a_+)$\quad (f.) & \quad ${(|A_+A_-a_+)}$\quad (f.)
\end{tabular}
\\
\hline
%----------------------------------------------
2\phantom{10}&
\begin{tabular}{l}
{$(|A_+A_-a_+)$}
\end{tabular}
 &
\begin{tabular}{p{60mm}|p{60mm}}
\quad $(a_+|A_+A_-a_+)$\quad (f.)  & \quad$(|A_+a_-a_+)$\quad (f.) 
\end{tabular}
\\
\hline
\hline
%----------------------------------------------

1\phantom{10}&
 \begin{tabular}{l}
{$(|A_+A_+a_-)$}
\end{tabular}
 &
\begin{tabular}{p{60mm}|p{60mm}}
\quad$(a_+|A_+A_+a_-)$\quad (f.) & \quad ${(|A_+A_+A_-)}$\quad (f.)
\end{tabular}
\\
\hline
%----------------------------------------------
2\phantom{10}&
\begin{tabular}{l}
{$(|A_+A_+A_-)$}
\end{tabular}
 &
\begin{tabular}{p{60mm}|p{60mm}}
\quad$(a_+|A_+A_+A_-)$\quad (f.)  & \quad$(|A_+A_+a_-)$\quad (f.) 
\end{tabular}
\\
\hline 
\hline

1\phantom{10}&
 \begin{tabular}{l}
{$(A_+|a_-A_+)$}
\end{tabular}
 &
\begin{tabular}{p{60mm}|p{60mm}}
\quad$(a_+A_+|a_-A_+a_+)$\quad (f.) & \quad ${(A_+|A_-A_+)}$\quad (f.)
\end{tabular}
\\
\hline
%----------------------------------------------
2\phantom{10}&
\begin{tabular}{l}
{$(A_+|A_-A_+)$}
\end{tabular}
 &
\begin{tabular}{p{60mm}|p{60mm}}
\quad $(a_+A_+|A_-A_+a_+)$\quad (f.)  & \quad$(A_+|a_-A_+)$\quad (f.) 
\end{tabular}
\\
\hline 

\end{tabular}
\caption{Bifurcations of branches of ILMs that  have the length of the code shorter than 4 but were not included in Tables \ref{Tab:a}-\ref{Tab:a+0a-}. The notation is the same as in Tables \ref{Tab:a+a-} and \ref{Tab:a+0a-}. Groups of codes linked through different types of bifurcations are separated.}
\label{Tab:Rest}
\end{table}
%\newpage

\end{document}